\documentclass[12pt]{article}
\pdfoutput=1
\usepackage{amssymb,amsmath,dsfont,verbatim}
\usepackage[normalem]{ulem}
\usepackage{slashed}
\usepackage{epsfig}
\usepackage{epstopdf}
\usepackage{latexsym}
\usepackage{graphicx}
\usepackage{booktabs}
\usepackage{bbm}
\usepackage{xspace}
\usepackage{cancel}
\usepackage[utf8]{inputenc}
\usepackage[babel]{csquotes}
\usepackage{enumitem}
\usepackage[numbers,compress]{natbib}
\usepackage[T1]{fontenc}
\usepackage[margin=20pt,small]{caption}
\usepackage{subfig}
\usepackage[toc]{appendix}
\usepackage{color}
\usepackage{float}
\usepackage{datetime}
\usepackage[
colorlinks=false,
linkcolor=darkblue,  
urlcolor=blue,    
filecolor=blue,     
citecolor=red,
linktocpage=true,
pdfstartview=FitV,
bookmarksopen=true,
hidelinks
]{hyperref}
\numberwithin{equation}{section}

\ifpdf
\fi
\newcommand*{\boxedcolor}{red}
\makeatletter
\renewcommand{\boxed}[1]{\textcolor{\boxedcolor}{%
		\fbox{\normalcolor\m@th$\displaystyle#1$}}}
\makeatother

	\newcommand{\ee}{\end{equation}}

\newcommand{\D}{\Delta}
\newcommand{\hD}{\widehat\D}
\newcommand{\Disp}{\text{D}}
\newcommand{\Flux}{\text{V}}
\newcommand{\Cd}{C_{\Disp}}
\newcommand{\hC}{\widehat{C}}

\newcommand{\htau}{\widehat{\tau}}

\newcommand{\hO}{\widehat{O}}

\definecolor{cardinal}{rgb}{0.6,0,0}
\definecolor{darkgreen}{rgb}{0,0.5,0}
\definecolor{golden}{rgb}{0.92, 0.7, 0}
\definecolor{midnight}{rgb}{0, 0, 0.5}
\definecolor{darkblue}{rgb}{0.2, 0, 0.8}

\def\sl{\frak{sl}}

\begin{document}  
	
	\begin{titlepage}
		
		\medskip
		\begin{center} 
			{\Large \bf Bootstrapping boundary-localized interactions II:

			Minimal models at the boundary}

			\bigskip
			\bigskip
			\bigskip
			
			{\bf Connor Behan$^1$, Lorenzo Di Pietro$^{2,3}$, Edoardo Lauria$^{4}$ and Balt C. van Rees$^4$\\ }
			\bigskip
			\bigskip
			${}^{1}$
			Mathematical Institute, University of Oxford, Andrew Wiles Building, Radcliffe Observatory Quarter, Woodstock Road, Oxford, OX2 6GG, UK\\
			${}^{2}$
			Dipartimento di Fisica, Universit\`{a} di Trieste, Strada Costiera 11, I-34151 Trieste, Italy\\
			${}^{3}$
			INFN, Sezione di Trieste, Via Valerio 2, I-34127 Trieste, Italy\\
			${}^{4}$
			CPHT, CNRS, Ecole Polytechnique, Institut Polytechnique de Paris, Route de Saclay, 91128 Palaiseau,
			France
			\vskip 5mm				
			\texttt{behan@maths.ox.ac.uk,~ldipietro@units.it,\\edoardo.lauria@polytechnique.edu,~balt.van-rees@polytechnique.edu} \\
		\end{center}
		
		\bigskip
		\bigskip
		
		\begin{abstract}
			\noindent 
			We provide evidence for the existence of non-trivial unitary conformal boundary conditions for a three-dimensional free scalar field, which can be obtained via a coupling to the $m$'th unitary diagonal minimal model. For large $m$ we can demonstrate the existence of the fixed point perturbatively, and for smaller values we use the numerical conformal bootstrap to obtain a sharp kink that smoothly matches onto the perturbative predictions. The wider numerical analysis also yields universal bounds for the spectrum of any other boundary condition for the free scalar field. A second kink in these bounds hints at a second class of non-standard boundary conditions, as yet unidentified.
		\end{abstract}

		\noindent

	\end{titlepage}
	
	
	\setcounter{tocdepth}{2}	
	
	\tableofcontents
	\newpage
	
\section{Introduction and summary}
\label{sec:intro}
Conformal boundary conditions pose a challenging classification problem: what is the set of possible boundary conformal field theories (BCFT) for a given bulk CFT? In this paper we continue the investigation  of this problem initiated in \cite{Behan:2020nsf} for the simplest possible bulk conformal field theory: a free massless scalar. Even in this simple case there is a rich set of possibilities to explore when we go beyond the well-known Neumann and Dirichlet `free' boundary conditions and we allow for interactions with boundary degrees of freedom, see e.g. \cite{Herzog:2017xha, Giombi:2019enr, DiPietro:2020fya}.

In the previous work \cite{Behan:2020nsf} we showed how to phrase the question in a way that makes it amenable to be attacked using the numerical conformal bootstrap \cite{Rattazzi:2008pe} (see also the review \cite{Poland:2018epd} and \cite{Liendo:2012hy,Gliozzi:2015qsa} for previous approaches to the bootstrap of BCFTs). This allowed us to put rigorous bounds on the space of conformal boundary conditions for a free scalar in a 4d bulk. The goal of the present work is to apply the same technique in the case of a 3d bulk. We recall that our method relies on three universal properties of the boundary conditions for a free field: (i) the existence of two operators $\hO_1\sim\phi$ and $\hO_2\sim \partial_\perp \phi$ with protected scaling dimensions and corresponding to the boundary modes of the bulk field,  (ii) a relation between the OPE coefficients of $\hO_1$ and $\hO_2$ with any other boundary operator, descending from the consistency with the (exactly known) bulk OPE expansion and (iii) the existence of special boundary operators with protected scaling dimensions coming from the bulk higher-spin currents. All these properties can be implemented in the bootstrap approach when studying the crossing symmetry constraints on the mixed set of boundary four-point functions of $\hO_1$ and $\hO_2$.

A convenient parametrization of the space of boundary conditions is in terms of the spin 2 gap $\hD_2$, which measures the non-locality of the boundary, and of the parameter $a_{\phi^2}$, i.e. the coefficient of the one-point function of the operator $\phi^2$. The latter also encodes the coefficients of $\hO_1$ and $\hO_2$ in the bulk-to-boundary OPE of $\phi$, and enters in the relation between boundary OPE coefficients mentioned above. Moreover, a novel feature in the 3d/2d setup is that the unitarity bound for boundary scalar operators is 0. To isolate the contribution of the identity operator one is then forced to also introduce a scalar gap $\hD_0 > 0$. Therefore we obtain our numerical bounds in the space of the three parameters $(a_{\phi^2},\hD_2, \hD_0)$ (see figure \ref{fig:3dplot}).  

The study for a 4d bulk showed that indeed the bootstrap problem is constrained enough to carve out a large chunk of the parameter space of possible boundary conditions. It also revealed the existence of an enticing kink that is naturally conjectured to correspond to an interacting boundary condition, previously unknown. However that analysis was `shooting in the dark', namely there was no input from data or perturbative constructions to indicate what kind of interacting boundary conditions might exist, and where they might be in the space of parameters.\footnote{See \cite{DiPietro:2020fya} for attempts at constructing interacting boundary conditions for the 4d scalar. In particular an example is discussed in their appendix B, though one that lies well within the allowed region of \cite{Behan:2020nsf}.} 

The situation is different in the 3d setup in that a perturbative construction exists, and it is seen to lie at the edge of the numerical exclusion plots. This interacting boundary condition can be reached through a coupling of the bulk scalar with Dirichlet boundary condition to unitary diagonal minimal models $\mathcal{M}_{m,m+1}$ on the 2d boundary, via the Lagrangian
\begin{equation}\label{eq:mmlag}
g \partial_\perp \phi \,\Phi_{(1,2)} + h \Phi_{(1,3)}~,
\end{equation}
where $\Phi_{(p,q)}$ denote the primaries of the minimal model. The RG flow triggered by this deformation has a perturbative fixed point at large $m$ with $g_*,h_*\sim 1/m$. In the numerical bootstrap analysis we find a family of kinks whose observables are in good agreement with the perturbative prediction for large $m$ (see figure \ref{3d2dzoom1}). This allows us to follow the fixed point to strong coupling and to establish the existence of the interacting boundary condition all the way to $m=4$. Surprisingly, the agreement with the perturbative prediction remains good even for such a low value of $m$. The remaining case of $m=3$, i.e. the coupling to the 2d Ising model, can be seen as a special case of the Lagrangian studied in \cite{Behan:2017dwr, Behan:2017emf} and based on that we conjecture that it flows to the Neumann boundary condition. 

In addition to these `boundary minimal models', we find an additional more mysterious family of kinks for lower values of the scalar gap $\hD_0$ (see figures \ref{fixed-a}-\ref{kink-series}), for which we do not propose a microscopic construction. Another feature that we observe and for which we lack an explanation is a certain splitting of the minimal model kink beyond $a_{\phi^2}\gtrsim 0.36$, close to the Neumann boundary condition (see figure \ref{3d2dzoom2}). In fact, we were not able to rigorously establish in perturbation theory the existence of any interacting fixed point from a perturbation of Neumann in 3d. Both for the `boundary minimal models' and for the family of kinks for lower $\hD_0$ we use extremal functional methods to extract the low lying spectrum, finding agreement with the perturbative calculations in the first case. Moreover, in order to verify that the solution to crossing is indeed describing a BCFT rather than an arbitrary non-local theory, we verify that the lowest lying protected operators satisfy the Ward Identity that descends from their parent bulk higher-spin currents.

The rest of the paper is organized as follows. In section \ref{sec:review} we review the derivation of the universal properties of a free scalar BCFT. In section \ref{sec:examples} we discuss the perturbative construction of interacting boundary conditions, in particular we show the existence of the `boundary minimal model' fixed points. Section \ref{sec:numset} contains a brief description of the numerical setup, which is then used in section \ref{sec:numericalresults} to obtain our main results, namely the constraints from the numerical bootstrap. We conclude in section \ref{sec:conclusions} by mentioning some possible future directions. Various appendices contain technical results that we use along the way.

\section{Review on analytic constraints on the free scalar BCFT}
\label{sec:review}
In this section we review the analytic constraints discussed in \cite{Behan:2020nsf} on the data that characterize the free scalar BCFT.  We consider a free massless scalar field $\phi$ in $d>2$ dimensions with a planar boundary.  We denote the  components of $x = (\vec{x},y)\in \mathbb{R}^{d-1} \times \mathbb{R}_+ $ as $x^\mu$, $\mu = 1,\dots, d$ where $x^d = y\geq 0$ is the direction orthogonal to the boundary at $y=0$, and those of $\vec{x}\in\mathbb{R}^{d-1}$ as $x^a$, $a=1,\dots, d-1$. We are interested in local and unitary boundary conditions that preserve the boundary conformal symmetry $SO(d,1)$.

\subsection{Protected boundary spectrum}
The spectrum of operators in the free scalar BCFT includes infinitely many boundary primaries with protected scaling dimensions. In any unitary free scalar BCFT this property follows from the bulk equation of motion. This section is a review of results that have already appeared in the BCFT literature. In particular, the spectrum of boundary modes of the free scalar was discussed in \cite{McAvity:1995zd,Liendo:2012hy,Dimofte:2012pd,Gaiotto:2014gha,Gliozzi:2015qsa,Billo:2016cpy,Giombi:2019enr}, the bulk-boundary crossing symmetry relations in eq.~\eqref{b1b2def} were derived in \cite{Liendo:2012hy,Gliozzi:2015qsa}, while higher-spin displacement operators were discussed in \cite{Giombi:2019enr,Behan:2020nsf}.

\subsubsection{Boundary modes of the free scalar}
Consider the bulk-boundary OPE (bOPE) of a scalar operator $\phi$ in a generic BCFT \cite{McAvity:1993ue,McAvity:1995zd,Liendo:2012hy}
\begin{align}
	\begin{split}\label{BOPEphi}
		\phi(\vec{x},y) 
		\underset{y\rightarrow 0}{\sim} \sum_i b_i~y^{\hD_i-\D_\phi}\,\widehat{{O}}_i (\vec{x})+\dots~.
	\end{split}
\end{align}
By rotational symmetry, the sum above is restricted to scalar boundary operators only. The ellipsis stands for boundary descendant operators, i.e. total derivatives with respect to the parallel coordinates. We will take unit-normalized boundary operators unless explicitly stated (see appendix~\ref{app:conventions} for a summary of our conventions). 

We now specify $\phi$ to be a free massless field. The Laplace equation dictates that $\Delta_\phi=\frac{d}{2}-1$ and, if $b_i\neq 0$, it also implies that the scaling dimension of $\hO_{i}$ can only take two values
\begin{align}
	\label{Deltai}
	\quad \widehat{\Delta}_1= \frac{d}{2}-1~, \quad \widehat{\Delta}_2 = \frac{d}{2}~.
\end{align}
The bulk-boundary couplings $b_i$, which are real numbers in unitary theories, are further constrained by the bulk-boundary crossing symmetry of the two-point correlation function of $\phi$. 
Taking $\phi$ to be unit-normalized, the bulk-boundary couplings $b_i$ satisfy the following relation \cite{Liendo:2012hy,Gliozzi:2015qsa}
\begin{align}\label{b1b2def}
	b_1^2=1+2^{d-2}a_{\phi^2}~,\quad b_2^2=\left(d-2\right)(1-2^{d-2}a_{\phi^2})~,
\end{align}
where $a_{\phi^2}$ is defined as
\begin{align}\label{one_point}
	\langle \phi^2(\vec{x},y)\rangle = \frac{a_{\phi^2}}{y^{d-2}}~.
\end{align}
Therefore unitarity restricts $a_{\phi^2}$ to lie in an interval
\begin{align}\label{aphi2_range}
	-\frac{1}{2^{d-2}} = a^{(D)}_{\phi^2} \leq a_{\phi^2}\leq a^{(N)}_{\phi^2} = \frac{1}{2^{d-2}}.
\end{align}
As we indicated above, the boundaries of the interval correspond to the Dirichlet ($b_1 = 0$) and Neumann ($b_2 = 0$) boundary condition.

\subsubsection{Higher-spin displacement operators}
The $\phi\times\phi$ OPE contains, together with the bulk stress-tensor $T\equiv J _2$, infinitely many higher-spin conserved currents $J_\ell$ with even spin $\ell\geq 4$. The conservation of these currents is generically violated by boundary-localized terms, so that
\begin{align}\label{ward_currents}
	\langle\partial_{\mu} J_\ell^{\mu \mu_1\dots \mu_{\ell-1}}(\vec{x},y)\dots\rangle&=\delta (y)\langle\widehat{\mathcal{O}}_\ell^{\mu_1\dots \mu_{\ell-1}}(\vec{x})\dots\rangle~.
\end{align}
Note that indices on the r.h.s are labels when they take the value $d$. Therefore, the BCFT generically contains boundary operators $\Disp_\ell^{(l)}$ and $\Flux_\ell^{(l+1)}$ of spin $l$ and $l+1$, respectively, and protected dimensions $\hD=d+\ell-2$, where $l$ is an even integer ranging from $0$ to $\ell-2$.  It is important to remark that the appearence of such protected boundary primaries hinges on the existence of the higher-spin conserved currents of the free scalar CFT.
In the case of the bulk stress-tensor, the Ward identity in eq.~\eqref{ward_currents} with $\ell=2$ implies that the boundary spectrum contains an $l=0$ primary with scaling dimension $\hD=d$. This operator, which controls the breaking of translations in the direction transverse to the boundary, is the so called \emph{displacement} operator and will be denoted as $\Disp \equiv \Disp_{2}^{(0)}$. In the same boundary multiplet, generically we expect a vector operator with $l=1$ and scaling dimension $\hD=d$ which controls the breaking of translations in the directions parallel to the boundary. This is the so called \emph{flux operator}, denoted as $\Flux^{(1)} \equiv \Flux_2^{(1)}$. As shown in \cite{Behan:2020nsf}, in any unitary BCFT where the parallel translations are preserved, locality implies that the flux operator is absent, i.e.
\begin{equation} \label{eq:loccond}
	\text{V}^{(1)} = 0~.
\end{equation}

\subsection{Exact OPE relations and bulk-boundary crossing}
The three-point correlation functions between the boundary modes of $\phi$ and a generic boundary primary satisfy certain exact OPE relations that ensure analyticity of the $\phi\times \phi$ OPE. Bulk-boundary crossing symmetry imposes further relations between the BCFT data of the protected boundary operators. Taken together these constraints will play a crucial role in our numerical bootstrap analysis.

\subsubsection{Constraints from analiticity of the bulk OPE}
Consider the three-point correlation function between the operators $\hO_i$ (the boundary modes of the free scalar) and a generic boundary primary $\widehat{\mathcal{O}}^{(l)}$ of scaling dimension $\hD$  transforming as a symmetric and traceless $SO(d-1)$ tensor of spin $l$. In any conformally invariant boundary condition these correlators have the following form \cite{Osborn:1993cr,Costa:2011mg}
\begin{align}\label{3ptssl}
	\langle \widehat{O}_i (\vec{x}_1)\widehat{O}_j (\vec{x}_2){\widehat{\mathcal{O}}^{(l)}} (\theta,\infty)\rangle=&\,\frac{\hat{f}_{ij{\widehat{\mathcal{O}}^{(l)}}}}{|\vec{x}_{12}|^{\widehat{\Delta}_i+\widehat{\Delta}_j-\widehat{\Delta}}}\left(- \frac{\vec{x}_{12}\cdot {\theta}}{|\vec{x}_{12}|}\right)^l~.
\end{align}
Tensor indices are contracted by means of a polarization vector $\theta^a$, which is null $\theta \cdot \theta =0$ as allowed by tracelessness of $\widehat{\mathcal{O}}^{(l)}$. Bose symmetry requires that
\begin{align}\label{Bosesymm}
	\hat{f}_{ij{\widehat{\mathcal{O}}^{(l)}}}=(-1)^l {}\hat{f}_{ji{\widehat{\mathcal{O}}^{(l)}}}~,
\end{align}
therefore only even spins $l$ are allowed in \eqref{3ptssl} if $i=j$. Analiticity of the $\phi\times \phi$ OPE requires the following exact relations between the OPE coefficients \cite{Behan:2020nsf}
\begin{align}
	\begin{split}\label{ssspinconstr}
		{}\hat{f}_{11{\widehat{\mathcal{O}}^{(l)}}}& =\kappa_1(\widehat{\Delta},l)\hat{f}_{12{\widehat{\mathcal{O}}^{(l)}}},\quad \kappa_1(\widehat{\Delta},l)\equiv -\frac{b_2 \Gamma \left(\frac{l+\widehat{\Delta}}{2}\right) \Gamma \left(\frac{d+l-\widehat{\Delta}-2}{2} \right)}{2b_1 \Gamma\left(\frac{d+l-\widehat{\Delta}-1}{2} \right)\Gamma \left(\frac{l+\widehat{\Delta}+1}{2}\right)}~,\\
		\hat{f}_{22{\widehat{\mathcal{O}}^{(l)}}}& =\kappa_2(\widehat{\Delta},l) \hat{f}_{12{\widehat{\mathcal{O}}^{(l)}}},\quad \kappa_2(\widehat{\Delta},l)\equiv -\frac{2 b_1 \Gamma \left(\frac{l+\widehat{\Delta}}{2}\right) \Gamma \left(\frac{d+l-\widehat{\Delta}}{2} \right)}{b_2\Gamma\left(\frac{d+l-\widehat{\Delta}-1}{2} \right)\Gamma \left(\frac{l+\widehat{\Delta}-1}{2} \right)}~.
	\end{split}
\end{align}
When $\mathcal{O}^{(l)}$ is one of the protected boundary operators among $\Disp_\ell^{(l)}$ or $\Flux_\ell^{(l+1)}$, some of the gamma functions appearing in eq.~\eqref{ssspinconstr} are singular and the relations above become degenerate. In particular we have to discriminate between two cases. If $\mathcal{O}^{(l)}$ is of the $\Disp_\ell^{(l)}$ type, then $\kappa_1=\kappa_2=\infty$ and so $\hat{f}_{12\Disp_\ell^{(l)}}=0$, leaving $\hat{f}_{11\Disp_\ell^{(l)}}$ and $\hat{f}_{22\Disp_\ell^{(l)}}$ unrelated to each other. We will see in the next subsection that exact relations of a different type are obeyed by $\hat{f}_{11\Disp_\ell^{(l)}}$ and $\hat{f}_{22\Disp_\ell^{(l)}}$. If $\mathcal{O}^{(l)}$ is of the $\Flux_\ell^{(l+1)}$ type, then $\kappa_1=\kappa_2=0$, so that $\hat{f}_{11\Flux_\ell^{(l+1)}}=\hat{f}_{22\Flux_\ell^{(l+1)}}=0$ and $\hat{f}_{12\Flux_\ell^{(l+1)}}$ is undetermined. For the $\Flux_\ell^{(l+1)}$ type of operators, the bulk-boundary crossing symmetry implies that whenever $\hat{f}_{12\Flux_\ell^{(l+1)}}\neq 0$ the operator must also appear in the bulk-boundary OPE of $J_\ell$, see e.g. appendix C of~\cite{Behan:2020nsf}.

\subsubsection{Constraints from bulk-boundary crossing}
\label{wardids}
We conclude this section by discussing some of the consequences of the bulk-boundary crossing symmetry for three-point functions involving $\Disp_\ell^{(0)}$ type of operators, and more specifically $\Disp$  and $\Disp_4^{(0)}$. We shall recall that a boundary operator $\Disp_\ell^{(0)}$ can appear in the bulk-to-boundary OPE of $\phi^2$, while it can couple to a bulk higher spin current $J_{\ell'}$ only if $\ell'=\ell$. Consequently the bulk-boundary crossing symmetry relates three-point function coefficients $\hat{f}_{ii\Disp_\ell^{(0)}}$ to the bulk-boundary OPE of $J_\ell$, as well as to the one-point function of $\phi^2$. For the displacement operator the precise relation was obtained in \cite{Behan:2020nsf} (see also appendix~\ref{app:dispWard}) and reads
\begin{align}\label{WardD20}
\hat{f}_{11\Disp}=\frac{(d-2) \left(a_{\phi^2} 2^d+2 C_{\Disp} S_d^2\right)}{4 (d-1) S_d b_1^2}, \quad 
\hat{f}_{22\Disp}=\frac{(d-2) \left(2 C_{\Disp} S_d^2-a_{\phi^2} 2^d\right)}{2 S_d b_2^2}.
\end{align}
In the equation above $S_d\equiv\text{Vol}(S^{d-1})={2 \pi ^{d/2}}/{\Gamma \left(d/2\right)}$ and $C_\Disp$ is the two-point correlation function of the displacement operator
\begin{align}\label{defCD}
	\langle \Disp(\vec{x})\Disp(0)\rangle = \frac{C_{\Disp}}{|\vec{x}|^{2d}}~.
\end{align}
By unitarity $C_{\Disp}\geq 0$.

Analogous relations should hold more generally for any $\Disp_\ell^{(l)}$, although deriving them may be more complicated. In the case of $\Disp_4^{(0)}$ we find (see appendix~\ref{app:D4Ward} for a derivation)
\begin{align}
\begin{split}
\label{WardD40}
	\hat{f}_{11\Disp_4^{(0)}}=& \frac{3 a_{\phi^2} 2^{d-4} (d-2) d^2}{b_1^2(d+1) (d+3) S_d}+\frac{(d-2) S_d\widehat{C}_{\Disp_4^{(0)}}}{8 b_1^2(d^2-1)}~,\\
\hat{f}_{22\Disp_4^{(0)}}=& -\frac{3 a_{\phi^2} 2^{d-2} (d-2) d^2}{b_2^2(d+3) S_d}+\frac{(d-2) S_d\widehat{C}_{\Disp_4^{(0)}}}{2 b_2^2(d-1)}~.
\end{split}
\end{align}
In the equation above we introduced a new quantity, $\widehat{C}_{\Disp_4^{(0)}}$, which is the two-point correlation function of the $\Disp_4^{(0)}$ operator, i.e.
\begin{align}\label{D4twoptdef}
	\langle \Disp_4^{(0)}(\vec{x}) \Disp_4^{(0)}(0) \rangle = \frac{\widehat{C}_{\Disp_4^{(0)}}}{|\vec{x}|^{2d+4}}~.
\end{align}
Unitarity requires $\widehat{C}_{\Disp_4^{(0)}}\geq 0$.

\section{Perturbative constructions in 3d/2d}
\label{sec:examples}

In this section we will discuss some perturbative constructions of interacting boundary conditions for the free massles scalar. These constructions are based on finding a short RG flow starting from either the Neumann or the Dirichlet boundary conditions and ending on a perturbative, unitary and interacting fixed point. We will start by briefly reviewing the results of \cite{Behan:2020nsf} for deformations that are linear in the boundary mode of the scalar, in generic $d$. Then we will consider specific examples: the deformation of the Dirichlet boundary condition via coupling to minimal models $\mathcal{M}_{m, m+1}$ on the boundary, which becomes perturbative at large $m$, and the deformation of the Neumann boundary condition via coupling to a Dirac fermion on the boundary, which becomes perturbative in the $d = 4-\epsilon$ expansion. We will conclude with some considerations about more general non-linear deformations of the Neumann boundary condition.

\subsection{Linear deformations (review of \cite{Behan:2020nsf})}\label{subsec:linear}
Let us consider first deformations of Dirichlet that are linear in $\partial_y\phi$, in generic $d$ bulk dimensions. We add a decoupled CFT$_{d-1}$ and deform by coupling to a boundary primary  $\widehat{\chi}$ of scaling dimension $\widehat{\Delta}_{\widehat{\chi}}=\frac{d}{2}-1-\epsilon$ with $0< \epsilon \ll 1$
\begin{align}\label{4d3dexampleDirichlet}
	S_{\partial}^{(D)}=S_{\text{CFT}_{d-1}} +g\int_{y=0}\,\mathrm{d}^{d-1} \vec{x}~ \partial_y\phi \,\widehat{\chi}~.
\end{align}
The modified Dirichlet boundary condition reads
\begin{align}\label{modifD}
	\phi\vert_{y=0}=-g\,\widehat{\chi}~.
\end{align}
As usual in conformal perturbation theory, we assume that any strongly relevant deformation (i.e. relevant for $\epsilon=0$) is tuned to zero, and this is consistent because when we do so the associated $\beta$ functions in any massless regularization scheme vanish. In specific examples, like the one we will study later, there might be additional approximately marginal  couplings, i.e. couplings that become marginal when $\epsilon$ is 0, in which case one needs to study the coupled $\beta$ functions for all of them. For the purpose of this section, we will not specify the $\beta$ functions and simply assume the existence of a perturbative fixed point with $g^2\propto \epsilon$. The first-order formulae we derive are valid as long as $g$ is the only approximately marginal  coupling between the bulk and the boundary. 

The stress-tensor of the CFT$_{d-1}$ gets an anomalous dimension due to the coupling to the bulk and becomes the lowest dimensional spin 2 operator on the boundary, which we call the pseudo stress-tensor and denote with $\htau^{ab}$. The anomalous dimension of $\htau$ at the leading order is
\begin{align}\label{anomstressdD}
	\widehat{\Delta}_{\widehat\tau}(g) & = d-1 +\widehat{\gamma}^{(1)}_{\widehat\tau} g^2 +O(g^4) ~,\quad 
	\widehat{\gamma}^{(1)}_{\widehat\tau} =\frac{\Gamma\left(\frac{d}{2}+1\right)}{\pi^{\frac{d}{2}}(d+1)}\frac{C_{\widehat{\chi}}^{(0)}}{C_{\widehat\tau}^{(0)}}~,
\end{align}
with the constant $C_{\widehat\tau}^{(0)}$ being  the `central charge' of the CFT$_{d-1}$ when $g=0$ i.e. 
\begin{align}\label{eq:tautau}
\begin{split}
	\langle \widehat{\tau}_{ab}(\vec{x})\widehat{\tau}_{cd}(0)\rangle_{g=0} & ={C_{\widehat\tau}^{(0)}}{}\frac{I^{ab,cd}(\vec{x})}{|\vec{x}|^{2d-2}}~,\\
	I^{ab,cd}(\vec{x})\equiv \frac{1}{2}[I^{ac}(\vec{x})I^{bd}(\vec{x}) & +I^{ad}(\vec{x})I^{bc}(\vec{x})]-\frac{1}{d-1}\delta^{ab}\delta^{cd}~,
\end{split}
\end{align}
with $I^{ab}(\vec{x}) = \delta^{ab}-2 x^a x^b /|\vec{x}|^2$. We are adopting the normalization in which the central charge for a $d$-dimensional free scalar CFT is \cite{Osborn:1993cr}
\begin{align}\label{totalCfree}
	C_{\widehat\tau}^{(0)}= \frac{d-1}{d-2}\frac{1}{ S_{d-1}^2}~.
	\end{align}
The normalization for $\widehat{\chi}$ and $\partial_y \phi$ is taken to be
\begin{align}
\langle\widehat{\chi}(0) \widehat{\chi}(\infty)\rangle_{g=0} =C_{\widehat{\chi}}^{(0)}~,\quad	\langle \partial_y \phi (0,0)  \partial_y \phi (\infty,0)\rangle_{g=0} \equiv C_{\partial_y\phi}^{(0)}=\frac{\Gamma{\left(\frac{d}{2}\right)}}{\pi^{\frac{d}{2}}}~.
\end{align}
The leading correction to $a_{\phi^2}$ and $\Cd$ are
\begin{align}
\begin{split}\label{aphi2CD}
a_{\phi^2} =&- 2^{2-d}+\delta a_{\phi^2}^{(1)} g^2+O(g^3) ~,\quad \delta a_{\phi^2}^{(1)} =\frac{2^{4-d}\pi^{\frac{d}{2}}C_{\widehat{\chi}}^{(0)}}{\Gamma\left(\frac{d}{2}-1\right)}~,\\
	C_{\text{D}} =&  \frac{\Gamma\left(\frac{d}{2}\right)^2}{2\pi^d}+ \delta C^{(1)}_{\text{D}} g^2 +O(g^4)~, \quad \delta C^{(1)}_{\text{D}} = -\frac{(d-2)\Gamma\left(\frac{d}{2}\right)C_{\widehat{\chi}}^{(0)}}{\pi^{\frac{d}{2}}}~.
\end{split}
\end{align}

We now consider the analogous deformation of Neumann, this time linear in $\phi$. We couple to a primary  $\widehat{\chi}$ of the CFT$_{d-1}$ with scaling dimension $\widehat{\Delta}_{\widehat{\chi}}=\frac{d}{2}-\epsilon$ and $0< \epsilon \ll 1$ 
\begin{align}\label{4d3dexampleNeumann}
	S_{\partial}^{(N)}=S_{\text{CFT}_{d-1}} +g\int_{y=0}\,\mathrm{d}^{d-1} \vec{x}~ \phi \,\widehat{\chi}~.
\end{align}
The same comments that we made above about tuning strongly relevant couplings and about the possibility of additional approximately marginal  couplings apply here. In particular the deformation by $\phi^2$ is tuned to zero. The case $d=3$ is special in that the Neumann b.c. admits an additional $\mathbb{Z}_2$ even, approximately marginal  deformation, namely $\phi^4$, and we will comment on this special case later. 
The modified Neumann boundary condition reads
\begin{align}\label{modifN}
	\partial_y\phi\vert_{y=0}=g\,\widehat{\chi}~.
\end{align}
Assuming again the existence of a perturbative fixed point with $g^2\propto \epsilon$, the anomalous dimension of the pseudo stress-tensor $\hat{\tau}$ at leading order reads
\begin{align}\label{anomstressdN}
	\widehat{\Delta}_{\widehat\tau}(g) & = d-1 +\widehat{\gamma}^{(1)}_{\widehat\tau} g^2 +O(g^4) ~,\quad 
	\widehat{\gamma}^{(1)}_{\widehat\tau} =\frac{\Gamma\left(\frac{d}{2}-1\right)}{4\pi^{\frac{d}{2}}}\frac{d}{d+1}\frac{C_{\widehat{\chi}}^{(0)}}{C_{\widehat\tau}^{(0)}}~,
\end{align}
where we assumed the normalizations
\begin{align}\label{phinorm}
	\langle\widehat{\chi}(0) \widehat{\chi}(\infty)\rangle_{g=0} =C_{\widehat{\chi}}^{(0)}~,\quad	\langle \phi (0,0)   \phi (\infty,0)\rangle_{g=0}\equiv C_{\phi}^{(0)}= \frac{\Gamma{\left(\frac{d}{2}-1\right)}}{2\pi^{\frac{d}{2}}}~.
\end{align}
The leading correction to $a_{\phi^2}$ and $\Cd$ under this deformation are
\begin{align}
\begin{split}\label{aphi2CDN}
	a_{\phi^2} =& 2^{2-d}+\delta a_{\phi^2}^{(1)} g^2+O(g^3) ~,\quad \delta a_{\phi^2}^{(1)} =-\frac{2^{3-d}\pi^{\frac{d}{2}}C_{\widehat{\chi}}^{(0)}}{\Gamma\left(\frac{d}{2}\right)}~,\\
	C_{\text{D}} =&  \frac{\Gamma\left(\frac{d}{2}\right)^2}{2\pi^d}+ \delta C^{(1)}_{\text{D}} g^2 +O(g^4)~, \quad \delta C^{(1)}_{\text{D}} = -\frac{\Gamma\left(\frac{d}{2}\right)C_{\widehat{\chi}}^{(0)}}{\pi^{\frac{d}{2}}}~.
\end{split}
\end{align}
As in the Dirichlet case, our final formulae only depend on the normalization of $\widehat{\chi}$ and on $C_{\widehat\tau}^{(0)}$.

\subsection{Minimal models coupled to Dirichlet}\label{minimalbc}
We will now consider the coupling of the bulk 3d free scalar to a particular set of boundary degrees of freedom: the unitary diagonal minimal model $\mathcal{M}_{m,m+1}$. The coupling we consider is
\begin{align}\label{D_minimal}
	\delta S_{\partial}=h\int_{y=0}\,\mathrm{d}^{2} \vec{x}~ \Phi_{(1,3)}+g\int_{y=0}\,\mathrm{d}^{2} \vec{x}~ \Phi_{(1,2)}{\partial_y \phi}~,
\end{align}
where $\Phi_{(p,q)}(z,\bar{z})$ denotes the Virasoro primary with Kac labels $(p,q)$, which we assume unit normalized and whose scaling dimension we denote with $\Delta_{(p,q)}$. At large $m$ we have
\begin{align}
\begin{split}
\hD_{(1,2)} & =\frac{1}{2}-\frac{3}{2 m}+O\left(\frac{1}{m^2}\right)~,\\
\hD_{(1,3)} & =2-\frac{4}{m}+O\left(\frac{1}{m^2}\right)~.
\end{split}
\end{align}
As a result the couplings $h$ and $g$ are approximately marginal  and more precisely weakly relevant. Only the coupling $g$ here couples the bulk to the boundary, however for the consistency of the RG in conformal perturbation theory we are also forced to add the second weakly relevant deformation $h$. 

\subsubsection{Large $m$ fixed points}
\label{subsubsec:largem} 
Conformal perturbation theory gives the following result for the $\beta$ functions
\begin{align}\label{beta_minimal_D}
\begin{split}
	\beta_h =&-\frac{4}{m} h+\pi  h^2 C_{(1,3)(1,3)}^{(1,3)}+\pi  g^2 C_{\partial_y\phi}^{(0)} C_{(1,2)(1,2)}^{(1,3)}+O(h^3, h g^2)~,\\
	\beta_{g}=&-\frac{3}{2m} g+2\pi  h g C_{(1,2)(1,3)}^{(1,2)}+O(g^3, h^2 g)~.
\end{split}
\end{align}
 The quantities $C_{(n,m)(r,s)}^{(r',s')}$ are the three-point structure constants of the $\mathcal{M}_{m,m+1}$ minimal model at $m=\infty$, which in our conventions read
\begin{align}\label{structureC}
	C_{(1,2)(1,2)}^{(1,3)}=C_{(1,3)(1,2)}^{(1,2)}=C_{(1,2)(1,3)}^{(1,2)}=-\frac{\sqrt{3}}{2}+O\left(\frac{1}{m}\right),\,\quad 
	C_{(1,3)(1,3)}^{(1,3)}=-\frac{4}{\sqrt{3}}+O\left(\frac{1}{m}\right).
\end{align}
Note that we are only using minimal model primaries of the form $\Phi_{(1, q)}$ which close among themselves for any $m$ according to fusion rules. Within this subsector, there is a $\mathbb{Z}_2$ selection rule such that $\Phi_{(1, q)}$ is even if and only if $q$ is odd.\footnote{This is different from the $\mathbb{Z}_2$ symmetry of the modular invariant theory which exists for integer $m$ and assigns the parity $(-1)^m$ to $\Phi_{(1,2)}$ \cite{Ruelle:1998zu}.} The possible terms in the $\beta$ function are restricted by the fact that the interaction \eqref{D_minimal} preserves the diagonal between this selection rule and the bulk $\mathbb{Z}_2$ symmetry $\phi \to -\phi$.
Setting the $\beta$ functions to zero we find three families of non-trivial fixed points, beside the gaussian one ($h_*=g_*=0$):

\begin{itemize}
	\item A family of fixed points at
	\begin{align}
		h_* =-\frac{\sqrt{3}}{\pi m}~, \quad  g_*=0~.
	\end{align}
	In this case the RG only involves the boundary local degrees of  freedom and the bulk theory just remains a decoupled Dirichlet boundary condition. The new 2d CFT at the IR fixed point is the minimal model with $m\to m-1$, because the RG triggered by $h$ is the famous 2d RG flow between two consecutive minimal models \cite{Zamolodchikov:1987ti,Cardy:1989da}.
	\item Two families of fixed points related by the bulk $\mathbb{Z}_2$ symmetry at
	\begin{align}\label{minimal_D_fixed_points}
		h_* =-\frac{\sqrt{3}}{2\pi m}~, \quad  g_*=\pm \frac{{2}}{\sqrt{\pi}  m}~.
	\end{align}
	This is a genuine interacting boundary condition for the bulk free scalar.
\end{itemize} 

For each of the fixed points above we can compute the scaling dimensions of the classically marginal operators associated to $h$ and $g$. These are obtained from the eigenvalues of the matrix of derivatives of the $\beta$ functions with respect to $h$ and $g$, evaluated at the fixed points.
For the interacting conformal boundary conditions we find that out of the two relevant deformations in the UV one linear combination becomes irrelevant in the IR, while the other one stays relevant, namely
\begin{align}
\label{Deltadim}
	\hD_{\pm} = 2\pm\frac{\sqrt{6}}{m}+O(1/m^2)~.
\end{align}
At the leading order in the large $m$ expansion, the operator with dimension $\hD_-$ is the lightest boundary operator after $\partial_y \phi\rvert_{y=0}$.

\subsubsection{Some observables of the interacting b.c.}
Next, we obtain the anomalous dimension of the pseudo stress-tensor on the boundary as well as the leading correction to $a_{\phi^2}$. At the leading order in the $1/m$ expansion we can extract these quantities by simply plugging into equations~\eqref{anomstressdD} and \eqref{aphi2CD} the fixed point value for $g$ in eq.~\eqref{minimal_D_fixed_points}.
We also need the central charge of the $m$-th minimal model in the $m=\infty$ limit, which in our conventions is
\begin{align}
	\hC_{\htau}^{(0)}=\frac{1}{2\pi^2}+O\left(\frac{1}{m}\right)~.
\end{align}
Substituting these quantities we find
\begin{align}\label{gammatauvsa_minimal_D}
	\hD_{\htau} = 2 +\frac{3}{2 m^2}+O\left(\frac{1}{m^3}\right)~,\quad a_{\phi^2} = -\frac{1}{2}+ \frac{8}{m^2}+O\left(\frac{1}{m^3}\right)~.
\end{align}
As a consequence of the fact that $g_*$ is a real number all these results are compatible with unitarity. This suggests that we have indeed found a family of unitary interacting conformal boundary conditions for the 3d free scalar CFT. In section~\ref{sec:numericalresults} we will find compelling evidence from the numerical conformal bootstrap that this family survives in the strong coupling limit, much beyond the expected validity of $1/m$ perturbation theory, up until $m>3$. From the general results of conformal perturbation theory we can also easily obtain the first correction to $C_{\Disp}$ at this fixed point. Plugging the value of $g_*$ from eq.~\eqref{minimal_D_fixed_points} in eq.~\eqref{aphi2CD} we find
\begin{align}\label{DminimalcptFirstCdvsa}
	C_{\Disp}=\frac{1}{8 \pi^2 }-\frac{2}{\pi^2 m^2}+O\left(\frac{1}{m^3}\right)~.
\end{align}

\subsubsection{The special case of $m=3$}
The minimal model with $m=3$, i.e. $\mathcal{M}_{3,4}$, is the 2d Ising model. In this case the operator $\Phi_{(1,2)}$ is the spin operator $\sigma$ of the Ising model. As a result the coupling $g$ in eq. \eqref{D_minimal} has the same form as the coupling of the $d$-dimensional Ising model to a scalar generalized free field (GFF) of dimension $\frac{d+s}{2}$ considered in \cite{Behan:2017dwr, Behan:2017emf}, with parameters $d=2$ and $s=1$. It was shown in that paper that this deformation triggers an RG flow to a fixed point whose nature depends on the range of $s$: in the interval $\frac{d}{2} < s < s_*$, $s_*$ being the value of $s$ for which the deformation stops being relevant, it is an interacting non-local fixed point known as long-range Ising. The defect construction of this fixed point was addressed in \cite{Paulos:2015jfa}. On the other hand for $s\leq \frac{d}{2}$ the IR fixed point is again a scalar GFF but with dimension $\frac{d-s}{2}$. Since in our setup $s=\frac{d}{2}=1$, the prediction from \cite{Behan:2017dwr, Behan:2017emf} is that we flow to a boundary condition with a scalar GFF of dimension $\frac{1}{2}$, which we can interpret simply as the Neumann boundary condition.\footnote{Note that $s=\frac{d}{2}$ is precisely the point at which the line of the long-range interacting fixed points joins the line of (generalized) free fixed points, see figure 1 of \cite{Behan:2017dwr}. This is reflected in the existence of a classically marginal operator, namely the operator quartic in the generalized free field, which for us is simply the $\phi^4$ deformation of the Neumann boundary condition. A one-loop calculation shows that this deformation is marginally irrelevant.}

One might be skeptical of the applicability of this result to our case, given that our deformation \eqref{D_minimal} contains also the additional coupling $h$. However the need to consider a flow in a two-dimensional coupling space is an artifact of large $m$, because only in that case are two approximately marginal operators present. Here we are describing a different way to approach the same IR fixed point that is not based on approximately marginal operators and perturbative fixed points, but rather uses the consistency of the family of RG's with different values of $d$ and $s$. In fact for generic $d$ and $s$ there is just one strongly relevant deformation that triggers the RG, while a second $\mathbb{Z}_2$-even relevant deformation proportional to the energy operator $\epsilon$ can be safely tuned to zero.\footnote{The $\mathbb{Z}_2$ here refers to the unbroken diagonal $\mathbb{Z}_2$ that flips both the spin operator and the bulk scalar field.} This second deformation maps to the unique $\mathbb{Z}_2$-even relevant deformation of the IR Neumann fixed point, namely $\phi^2$. As a consistency check, also the fixed point that we find at large $m$ contains only one relevant deformation besides the protected operators coming from the boundary modes of the scalar field (which are $\mathbb{Z}_2$-odd when $m=3$). This can be seen in perturbation theory and will be later confirmed also by the numerical analysis.

Based on these considerations we are led to conjecture that the continuation to $m=3$ of the large-$m$ fixed point is the free Neumann boundary condition. As a sanity check, we can verify that the boundary RG flow between Dirichlet + Ising and Neumann is allowed by the monotonicity of the 3d/2d boundary central charge $b$ of \cite{Jensen:2015swa}, i.e. that 
\begin{equation}
\frac{1}{2} = c_{\text{Ising}} \geq b_{\text{N}} - b_{\text{D}}~.
\end{equation}
Note that this is a non-trivial check because clearly $b_{\text{N}} - b_{\text{D}}\geq 0$, given that it is always possible to flow from Neumann to Dirichlet. From \cite{Jensen:2015swa} we see that $b_{\text{N}} = - b_{\text{D}} = \frac{1}{16}$, so that the constraint is indeed satisfied along the proposed RG. 

\subsection{Mixed Yukawa theory}
We now consider an example of a deformation of the Neumann boundary condition to an interacting one. This example is under control when we take the bulk dimension to be $4-\epsilon$ and $\epsilon \ll 1$. As a result we cannot firmly establish that the interacting boundary condition will keep existing when $\epsilon$ is extrapolated to 1. Concretely, we consider a free scalar in $d=4-\epsilon$ bulk dimensions with Neumann b.c., coupled to a Dirac fermion $\psi$ on the boundary with action
\begin{equation}\label{mixed_yukawa}
	S=\int_{y\geq0} \,\mathrm{d}^{d-1}\vec{x}  \,\mathrm{d}y \, \frac{1}{2}(\partial\phi)^2+ \int_{y=0}\,\mathrm{d}^{d-1}  \vec{x}~\left[\bar{\psi}\slashed{\partial} \psi + g\,\phi\,\bar{\psi}{\psi}\right]~.
\end{equation}
This theory has a $\mathbb{Z}_2$ symmetry which is the diagonal unbroken subgroup between the bulk symmetry $\phi\to-\phi$ and the boundary symmetry that flips the bilinear $\bar{\psi}\psi \to -\bar{\psi}\psi$ (this is a reflection symmetry when $\epsilon = 0$, i.e. for a 3d boundary). Using this symmetry we can avoid considering the approximately marginal operator $\phi^3$, and as usual strongly relevant deformations are tuned to zero. The modified Neumann b.c. is $\partial_y \phi =g\, \bar{\psi}{\psi}$.

This theory was studied previously in \cite{Herzog:2017xha}, where the existence of a fixed point for small $\epsilon$ was established.  The one-loop $\beta$ function was found to be
\begin{align}
	\beta_g =-\frac{\epsilon}{2}g+\frac{2}{3\pi^2}g^3+O(g^4)~.
\end{align}
giving an interacting IR fixed point for
\begin{align}\label{mixedY_fixedPoint}
	(g_*)^2 =\frac{3\pi^2}{4}\epsilon~.
\end{align}
The anomalous dimension of the pseudo stress-tensor, as well as the one-loop correction to $a_{\phi^2}$ and  $\Cd$ at this fixed point can be then obtained from the general formulae~\eqref{anomstressdN} and \eqref{aphi2CDN}. We need the central charge for a single 3d Dirac fermion, and the normalization of the two-point correlation function of $\bar{\psi}\psi$ in the free theory with $\epsilon=0$. These read \cite{Osborn:1993cr}
\begin{align}
	C^{(0)}_{\widehat\tau} = \frac{3}{16\pi^2}~,\quad 
	C^{(0)}_{\bar{\psi}\psi} = \frac{1}{16\pi^2}~.
	\end{align}
Plugging in \eqref{anomstressdN} and \eqref{aphi2CDN} we get
\begin{align}\label{mixed_yukawa_pred}
\begin{split}
	\frac{a_{\phi^2}}{a_{\phi^2}^{(N)}}=1-\frac{3 \pi ^2 \epsilon }{32}+O\left(\epsilon ^2\right)~,\quad 	\widehat{\Delta}_{\htau}=3-\frac{19 \epsilon }{20}+O(\epsilon^2)~,\quad 
	 \frac{\Cd}{\Cd^{(N)}} =1-\frac{3 \pi ^2 \epsilon }{32}+O\left(\epsilon ^2\right)~,
\end{split}
\end{align}
where $\Cd^{(N)}=\frac{\Gamma\left(\frac{d}{2}\right)^2}{2\pi^d}$ and $a_{\phi^2}^{(N)}=2^{2-d}$.

\subsection{No-go theorem for deformations of Neumann}

The Neumann boundary condition for a 3d free scalar admits a $\mathbb{Z}_2$-even classically marginal boundary deformation, namely $\phi^4$. We can use this operator to prove that any approximately marginal deformation of the Neumann boundary condition of the generic form
\begin{equation}\label{eq:ndef}
g \int_{y=0} \,\mathrm{d}^{2}\vec{x}  \, \phi^n \hat{\mathcal{O}}~,
\end{equation}
with $\Delta_\mathcal{O}= 2- \frac{n}{2} + \epsilon$, $\epsilon\ll1$ and $n=2,3$, or linear combinations of such deformations, cannot lead to a unitary interacting boundary condition under perturbative control at small $\epsilon$. This is simply because any such coupling will also generate
\begin{equation}
\frac{\lambda}{4!} \int_{y=0} \,\mathrm{d}^{2}\vec{x}  \,\phi^4 ~,
\end{equation}
at leading order in perturbation theory. Reading off the quadratic term from \cite{PhysRevLett.29.917}, $\beta_\lambda$ has the structure 
\begin{equation}
\beta_\lambda = \frac{3\lambda^2}{4\pi} + C \,g^2 + {\text{cubic terms}}~,
\end{equation}
where $C>0$. Therefore the only perturbative real solution to $\beta_\lambda = 0$ is $\lambda = g = 0$ and the bulk and the boundary are necessarily decoupled. This leaves as the only possibility for a perturbative fixed point around the Neumann boundary condition the case of a linear deformation, i.e. $n=1$ in \eqref{eq:ndef}, for which some general results were proven in section \ref{subsec:linear}. For a linear deformation, $g$ will only enter $\beta_\lambda$ at some higher order in perturbation theory and therefore we cannot easily draw general conclusions about the existence of fixed points.

\section{Numerical setup}\label{sec:numset}
The numerical setup we consider is exactly as in \cite{Behan:2020nsf}, with the one exception that now the boundary dimension is 2 instead of 3. In this section we briefly recall the main features, leaving some details to appendix \ref{app:K123} and the rest to the references.

We choose five crossing equations for the correlation functions
\begin{equation}
	\langle \hO_1 \hO_1\hO_1\hO_1 \rangle, \qquad \langle \hO_1 \hO_1\hO_2\hO_2 \rangle, \qquad \langle \hO_2 \hO_2\hO_2\hO_2 \rangle
\end{equation}
and analyze them with well known numerical bootstrap methods. Compared to a standard multi-correlator problem the main distinctive features of these correlation functions are the OPE relations of equation \eqref{ssspinconstr}. They show that operators with a generic dimension can only appear simultaneously in all three of the $\hO_1\times \hO_1$, $\hO_1\times \hO_2$ and $\hO_2\times \hO_2$, with related OPE coefficients.

Exceptions to this rule only occur for special scaling dimensions where the gamma functions in equation \eqref{ssspinconstr} are singular; in that case the equations are still valid in the obvious sense that certain OPE coefficients must be identically zero. These special dimensions include the protected operators originating from the higher-spin bulk current but more generally they simply coincide with the double-twist dimensions for $\hO_1$ and $\hO_2$.

At an operational level, one crucial implication of the OPE relations is that odd spin operators $X$ can only have the special dimensions, because for those $f_{11X} = f_{22X} = 0$ by Bose symmetry, and only by tuning their scaling dimensions can we then avoid having to set $f_{12X} = 0$ by the OPE relations.

For the even spin channels the OPE relations for generic dimension are straightforwardly implemented (see \cite{Behan:2020nsf} for details) and we can then add all the additional operators corresponding to the special dimensions by hand.\footnote{Note however that in the results below we did not add the scalar of dimension 1, unless indicated, if the assumed gap in the scalar sector is higher than 1.} This procedure however significantly increases the necessary degree of the polynomial approximations, and in practice we therefore only implemented it for spins 0 and 2, leaving the structure of the higher-spin channels as in a standard multiple-correlator problem.

After carrying out the steps in appendix \ref{app:K123}, the numerical problem was attacked with \texttt{SDPB} v2 \cite{SimmonsDuffin:2015,Landry:2019} and \texttt{PyCFTBoot} \cite{Behan:2016}.

\section{Numerical results}
\label{sec:numericalresults}
In this section we discuss the results obtained from the conformal bootstrap. To structure the discussion let us begin with an overview plot shown in figure \ref{fig:3dplot}. We show the maximal gap in the spin 2 sector, denoted $\hat \Delta_2$, as a function of $a_{\phi^2}$ and the assumed gap $\hat \Delta_0$ in the scalar sector. The three different regions in the plot correspond to the following results:

\begin{figure}
	\centering
	\includegraphics[width=\textwidth]{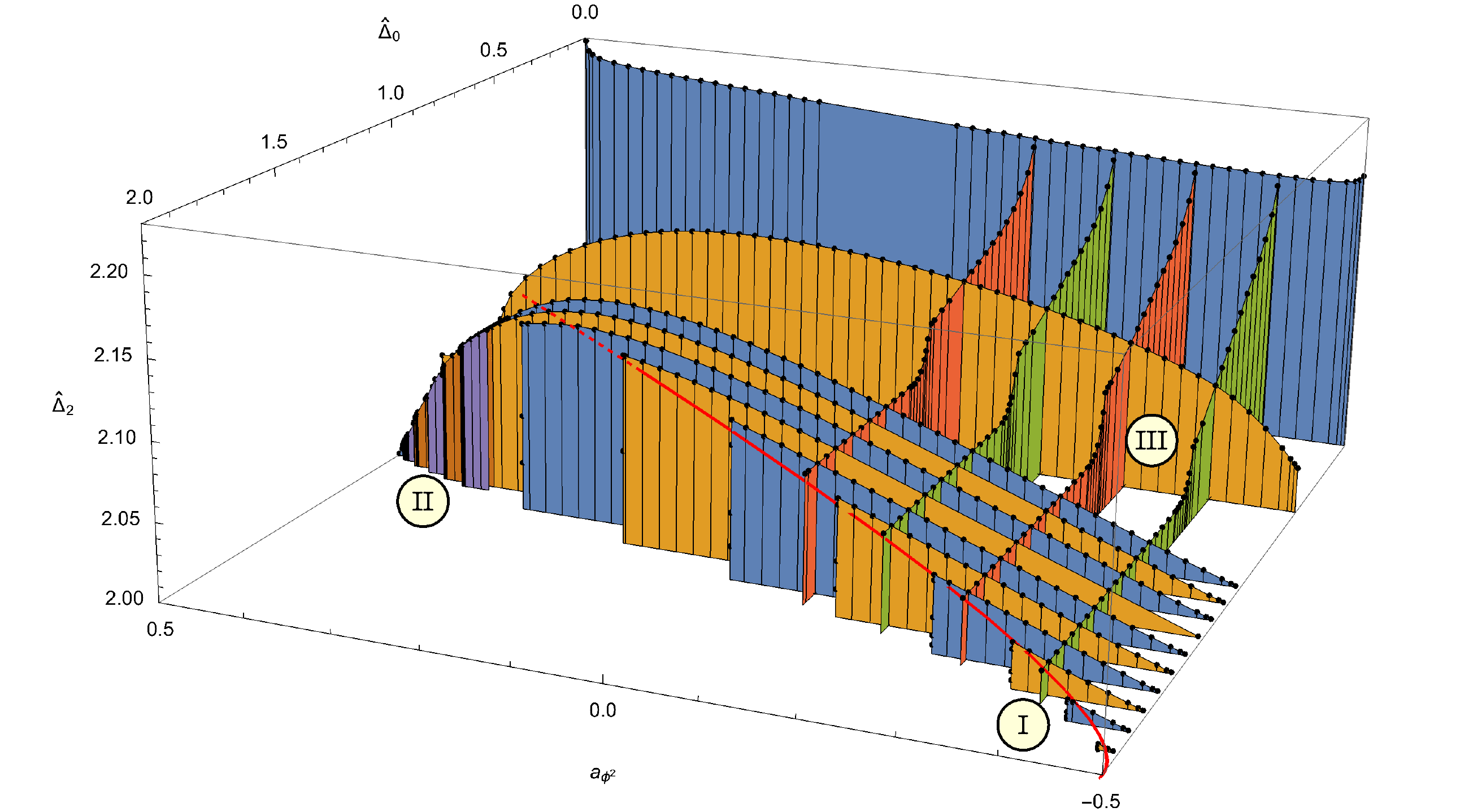}
	\caption{Overview plot of our main numerical results. We plot the maximal gap in the spin 2 sector as a function of $a_{\phi^2}$ and the assumed gap in the scalar sector. (Note the reflected axis for $a_{\phi^2}$ which decreases when moving to the right.) We find noteworthy structure in the three indicated regions. The red line is the perturbative result of subsection \ref{subsubsec:largem}.\label{fig:3dplot}}
\end{figure}

\begin{itemize}
	\item[I] In this region we can make contact with the perturbative analysis of subsection \ref{subsubsec:largem}, where we discussed coupling the $m$'th minimal model to the free boson with Dirichlet boundary conditions. This is best done by following the orange/blue slices; a more precise plot of these data points is shown in figure \ref{3d2dzoom1}. (In figure \ref{fig:3dplot} the perturbative result is already shown as a red line; the exact formulas are summarized below.)
	\item[II] Smaller values of $m$ correspond to larger values of $a_{\phi^2}$. Based on a crude extrapolation of the analytic results we expect that $m=4$ yields $a_{\phi^2} \approx 0$ approximately, and according to \cite{Behan:2017dwr,Behan:2017emf} the case with $m=3$ should correspond to $a_{\phi^2} = 1/2$ exactly. It is then interesting to search for signs of non-unitarity in the solution to the crossing equations with fractional $m$ between 3 and 4. This brings us into region II, a more detailed analysis of which is shown in figure \ref{3d2dzoom2}. The data for this plot  corresponds to the (poorly visible) purple and brown slices in figure \ref{fig:3dplot}.
	\item[III] For lower values of the scalar gap there is another noticeable but mysterious kink that we analyze further below, starting with figure \ref{fixed-a}. The outer envelopes in that figure correspond to the red and green slices of figure \ref{fig:3dplot}.
\end{itemize}

For completeness let us note that there is an intrinsic convexity property to the allowed region: for fixed $a_{\phi^2}$, if a point in  $(\hD_0, \hD_2)$ plane is excluded then so is every point with larger $\hD_0$ or larger $\hD_2$.

\subsection{Shrinking allowed regions}\label{ssec:spin2gap}

\subsubsection*{Region I}
\begin{figure}[h]
	\centering
	\includegraphics[width=0.85\textwidth]{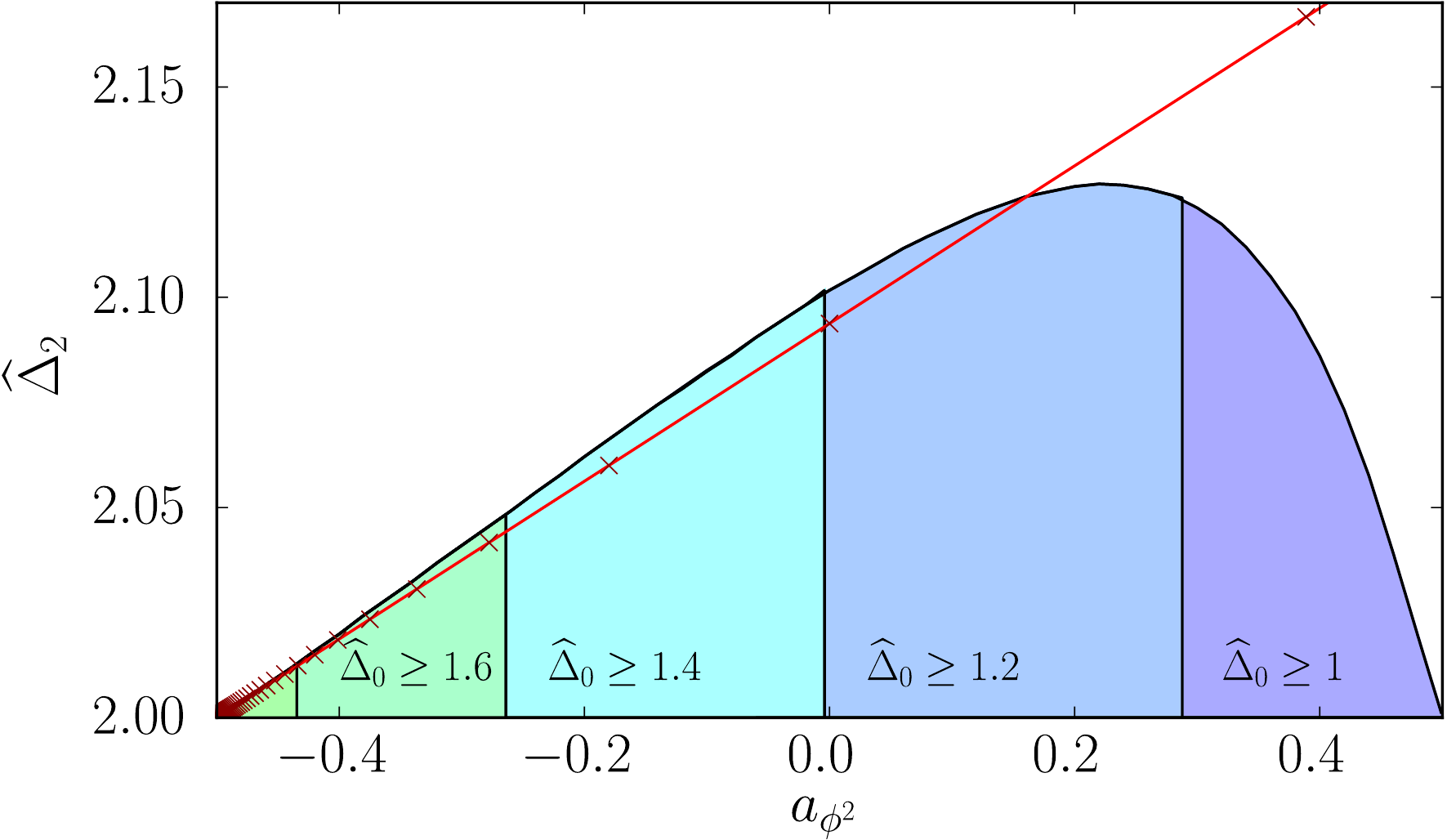}
        \caption{The maximum spin 2 gap for five choices of the imposed scalar gap. The bound is not smooth but contains a vertical line which sweeps to the left as the scalar gap is increased. The red crosses emanating from the left are the Dirichlet deformations of the $m$'th minimal model, eq.~\eqref{gammatauvsa_minimal_D}, extrapolated down to $m = 3$ (the rightmost value).}
	\label{3d2dzoom1}
\end{figure}

The blue slice furthest back in figure \ref{fig:3dplot} shows that the bound $\widehat{\Delta}_2$ on the first spin 2 operator is relatively weak if we set $\widehat{\Delta}_0 = 0$, so if we do not assume a gap for the first non-trivial scalar operator. Fortunately the bound however comes down quickly if we increase $\widehat{\Delta}_0$. For $\widehat{\Delta}_0 = 1$ we find that $\widehat{\Delta}_2 = 2$, the unitarity bound, both at the Dirichlet ($a_{\phi^2} = -1/2$) and the Neumann ($a_{\phi^2} = +1/2$) points, whereas for intermediate values of $a_{\phi^2}$ the bound is the outer envelope of figure \ref{3d2dzoom1}. Note that saturation at the endpoints does not mean that these points necessarily have a stress-tensor at the boundary: instead the double-twist vector $V^{(1)} = [\widehat{O}_1 \widehat{O}_2]_{0, 1}$ mimics a spin 2 primary of dimension 2 through the fake primary effect \cite{Karateev:2019}.\footnote{In fact, both for Dirichlet and Neumann boundary conditions one of the two operators whose correlation functions we analyze is actually set to zero. For the purpose of finding solutions to the crossing equations we can replace it with a decoupled generalized free field; this obeys the OPE relations.}
\begin{figure}
	\centering
	\includegraphics[width=0.85\textwidth]{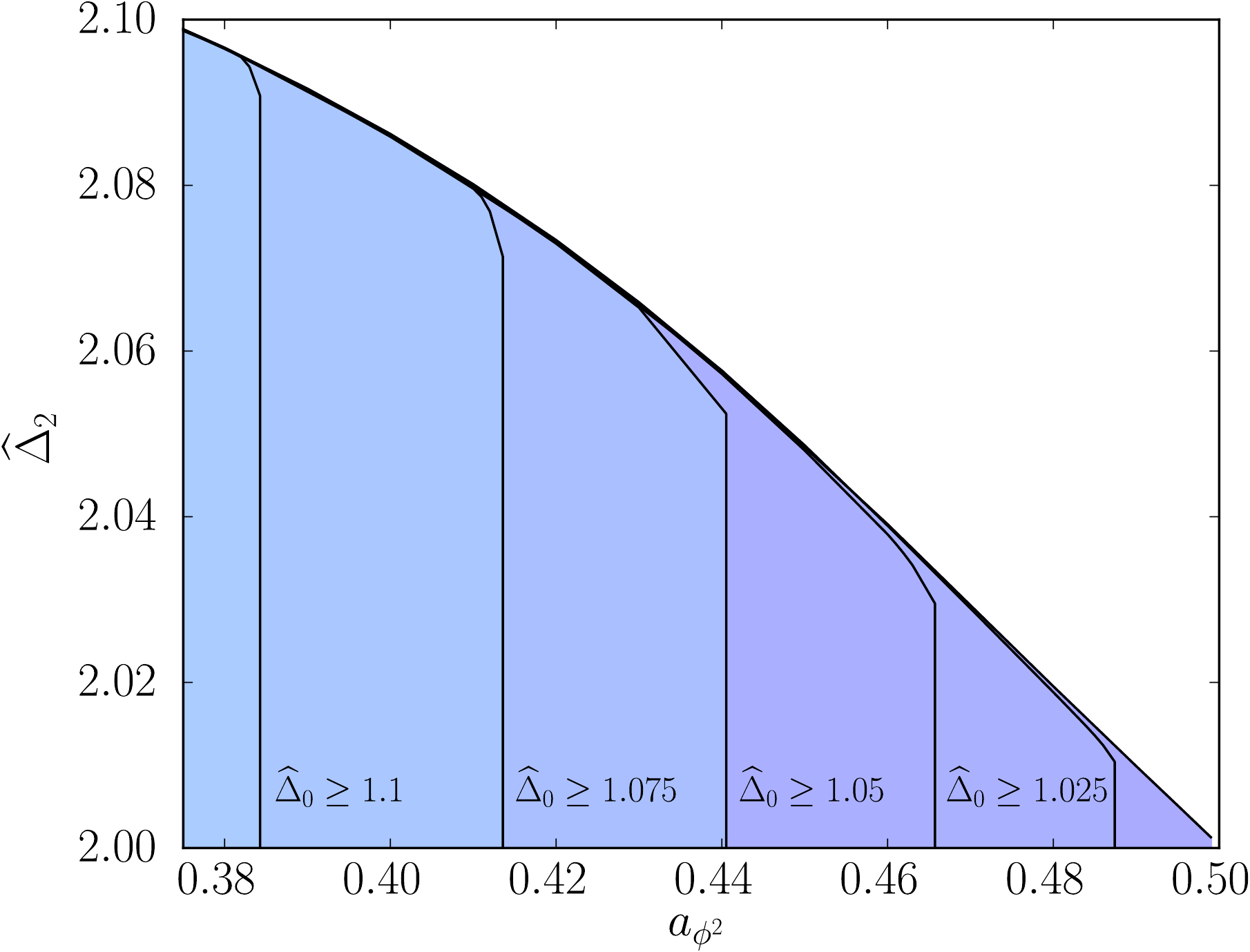}
	\caption{The maximum spin 2 gap, as in the previous figure, except with scalar gaps of $1.025$, $1.05$, $1.075$, $1.1$ and $1.125$. The upper bound on $\widehat{\Delta}_2$ within one of these regions is no longer the same as the upper bound for the $\widehat{\Delta}_0 = 1$ region. While it is possible that the upper right corners still correspond to minimal model boundary conditions, it is curious that this effect occurs in the part of the plot where non-unitarity is expected to be most pronounced.}
	\label{3d2dzoom2}
\end{figure}
Figure \ref{3d2dzoom1} also shows some of the slices for several other values with $\hD_0 \geq 1$. For any $a_{\phi^2}$ we see that the maximal spin 2 gap as a function of $\hD_0$ remains constant for some time, until it suddenly drops to the unitarity bound. For a wide range of values of $a_{\phi^2}$ the tip of the resulting vertical cliff in the three-dimensional space, which is clearly visible in figure \ref{fig:3dplot}, is in very good agreement with the perturbative estimates for the minimal model boundary conditions, which we recall were given by
\begin{equation}
\begin{split}
	a_{\phi^2}(m) &= - \frac{1}{2} + \frac{8}{m^2} + \ldots\\
	\hD_0(m) &= 2-\frac{\sqrt{6}}{m} + \ldots\\
	\hD_2(m) &= 2 + \frac{3}{2 m^2} + \ldots~,
\end{split}
\end{equation}
from equations \eqref{Deltadim} and \eqref{gammatauvsa_minimal_D}. For specific values of $m$ we also include the prediction for the spin 2 gap in the two-dimensional projection in figure \ref{3d2dzoom1}. Remarkably the prediction from conformal perturbation theory works well all the way down to $m = 4$. For $m=3$ (so the Ising model) our numerics exclude the perturbative prediction, but the latter is unlikely to be trustworthy in the first place.

\subsubsection*{Region II}
For $m=4$ the triplet of OPE data shown in the previous figures is in reasonable agreement with the location of the cliff, and so we expect this minimal model boundary condition to saturate our bounds near $a_{\phi^2} \approx 0$. What happens for lower $m$?

Let us first discuss $m = 3$. Our perturbative prediction for $m=3$ is that $a_{\phi^2} \approx 0.39$, but also that $\hD_0 \approx 1.18$ and that $\hD_2 \approx 2.17$, a triplet of OPE data which is easily excluded by the numerics. So we cannot trust extrapolated perturbation theory. As we mentioned above, it is in fact likely that the $m=3$ case flows to the Neumann point at $a_{\phi^2} = 1/2$ because of the connection to the long-range Ising model \cite{Behan:2017dwr,Behan:2017emf}, and our numerics certainly allow for this possibility.

Besides the fate of the $m=3$ minimal model boundary condition we may also inquire about the non-physical cases with $3 < m < 4$. Presumably the solution to the crossing equations continues to exist in this region, since we see no reason not to expect a reasonably smooth function of $m$, but one might expect it not to be consistent with the unitarity conditions. In equations, one expects that some of the coefficients multiplying the conformal blocks become negative. (Of course the same can be said about the other non-integer values of $m$, but the effect of these non-unitarities is probably most distinct for the lowest values of $m$.)

The above arguments provide sufficient motivation for a closer look at the regions which we expect to correspond to $3 < m \lesssim 4$. The most promising region for this search is region II in figure \ref{fig:3dplot}. The data for this region is also shown in figure \ref{3d2dzoom2}.

Interestingly, we do observe a qualitative change in this region: for $a_{\phi^2} \gtrsim 0.36$ the endpoint of the cliff (in the spin 2 gap) for a fixed $\hD_0 > 1$ starts to deviate from the curve given by $\hD_0 = 1$. In other words, in this region we find \emph{two} natural extremal points: one can be obtained by first extremizing the spin 2 gap and then the scalar gap, and the other by performing these extremizations in the reverse order. We are not entirely sure about the meaning of this phenomenon. Clearly the split happens close to the perturbative prediction for $a_{\phi^2}$ at $m=3$, but that might just be a numerical coincidence. We will return to this region when we discuss extremal spectra in the next subsection.

\begin{figure}
	\centering
	\subfloat[$n_{\mathrm{max}} = 4,5,6,7,8$]{\includegraphics[width=0.45\textwidth]{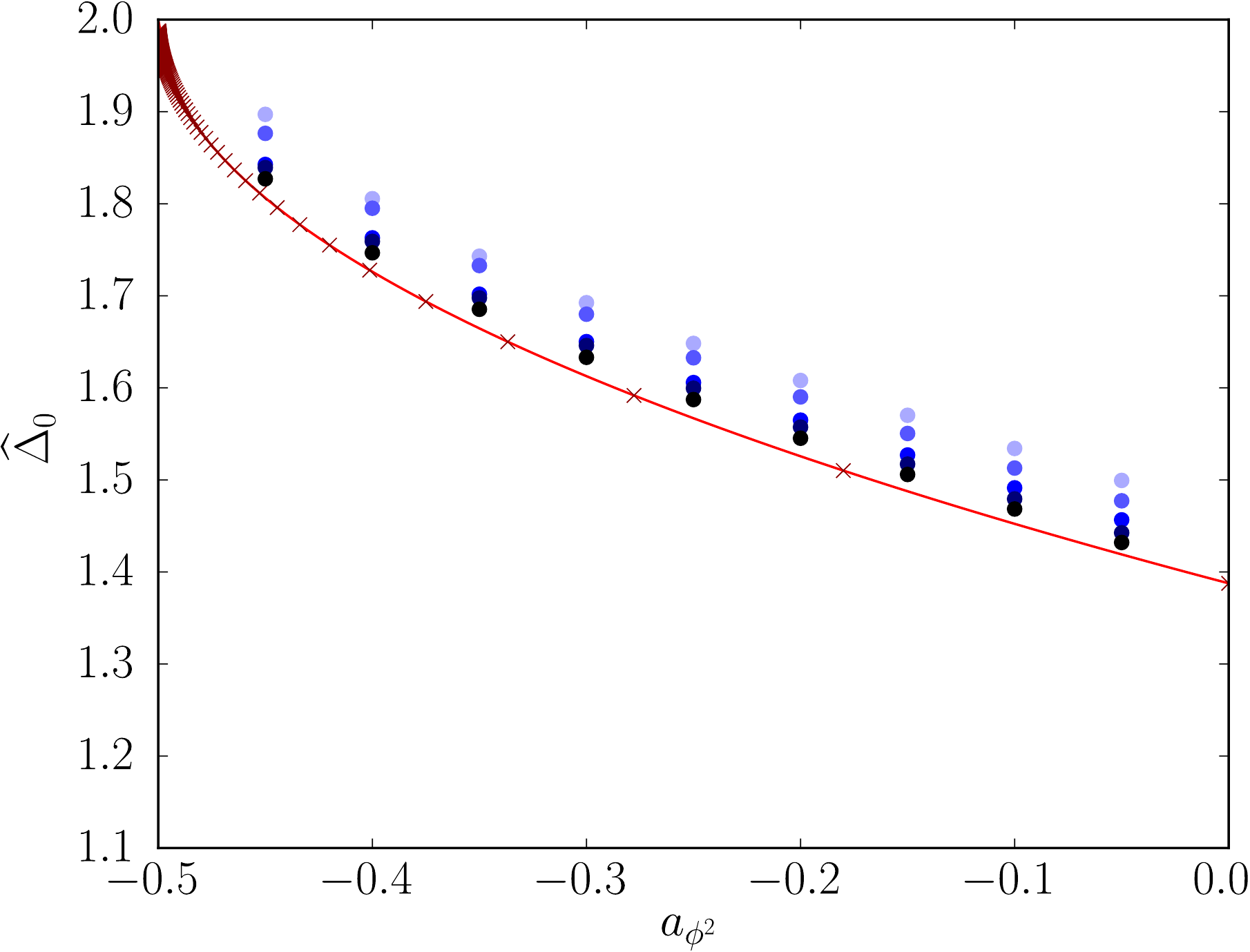}}
	\subfloat[$n_{\mathrm{max}} = 8$]{\includegraphics[width=0.45\textwidth]{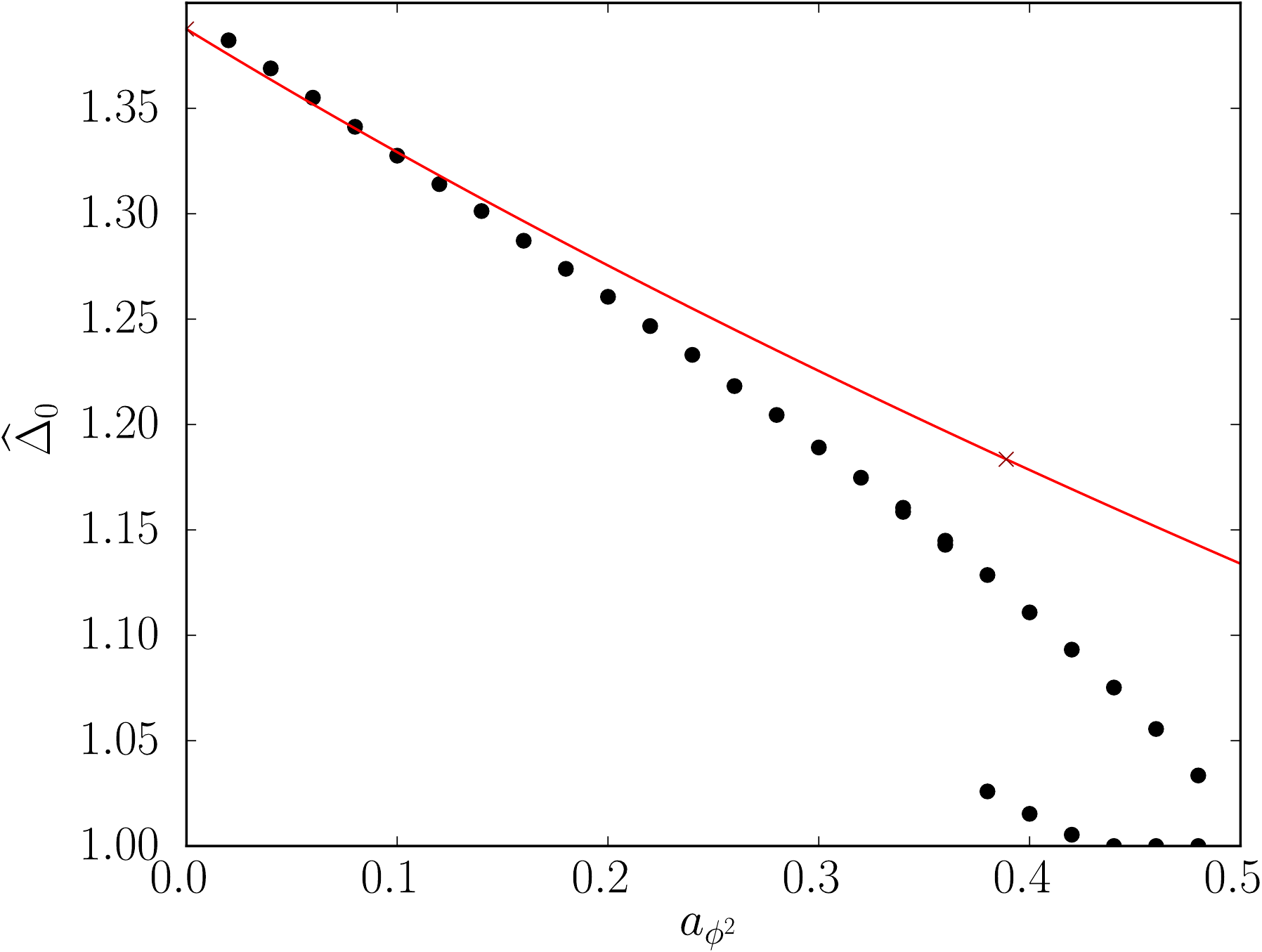}}
	\caption{The maximal gap for unprotected scalars plotted as a function of $a_{\phi^2}$. On the left, we focus on region I, which we take as $a_{\phi^2} < 0$, where conformal perturbation theory works best. The agreement between \eqref{minimal_D_fixed_points} and our data improves as the number of derivatives is increased. On the right, we have extended the plot to $a_{\phi^2} > 0$ but only for $n_{\mathrm{max}} = 8$ (black dots), the number of derivatives used everywhere else in this paper. For most values of $a_{\phi^2}$, the maximal scalar gap is the same whether or not we make assumptions about operators with spin 2. After reaching $a_{\phi^2} \approx 0.36$, however, this is no longer the case. The upper set of points here comes from maximizing $\hD_0$ with no assumptions while the lower set of points comes from maximizing $\hD_0$ along the line having maximal $\hD_2$.}
	\label{corner}
\end{figure}

The phenomenon discussed above can also be seen in figure \ref{corner} where we plot just the maximal scalar gap. For $a_{\phi^2} < 0$, which is displayed on the left, we not only see a good match with the perturbative prediction but also find that the match \emph{improves} by increasing $n_{max}$. This strengthens our conviction that the the minimal model boundary conditions indeed saturate our bounds. On the right of figure \ref{corner} we see an increasing mismatch in the region where perturbation theory is less trustworthy. In this plot we have also added a second set of data points, which correspond to the maximal dimension of the scalar if we first maximize the spin 2 gap --- in other words, if we move along the envelope of figure \ref{fig:3dplot}. The clear jump at about $a_{\phi^2} \approx 0.36$ is our best indication that something special is going on at that value.

\begin{figure}
	\centering
	\subfloat[$a_{\phi^2} = -0.1$, $\widehat{O}_1^2$ disallowed]{\includegraphics[width=0.45\textwidth]{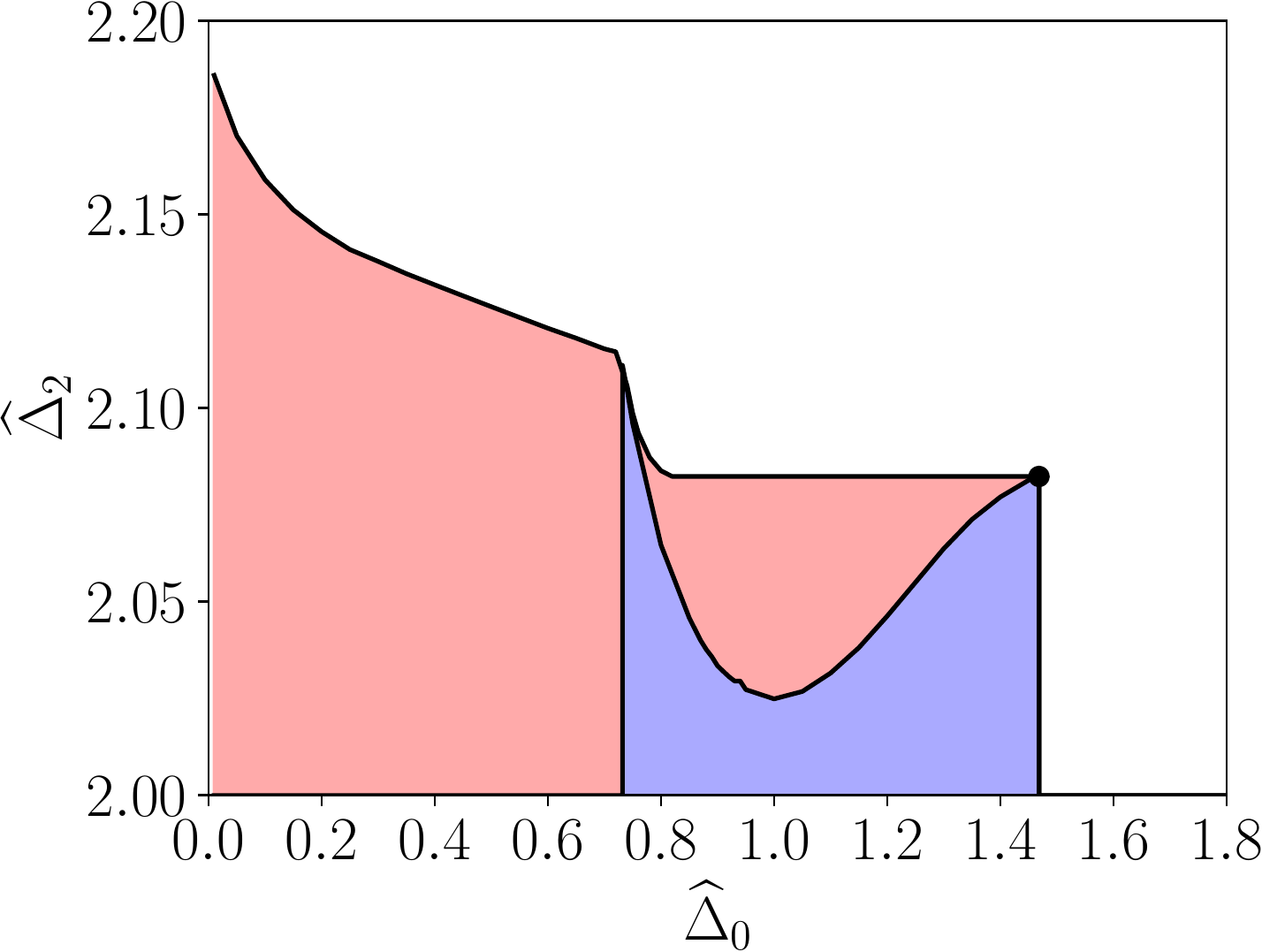}}
	\subfloat[$a_{\phi^2} = -0.4$, $\widehat{O}_1^2$ disallowed]{\includegraphics[width=0.45\textwidth]{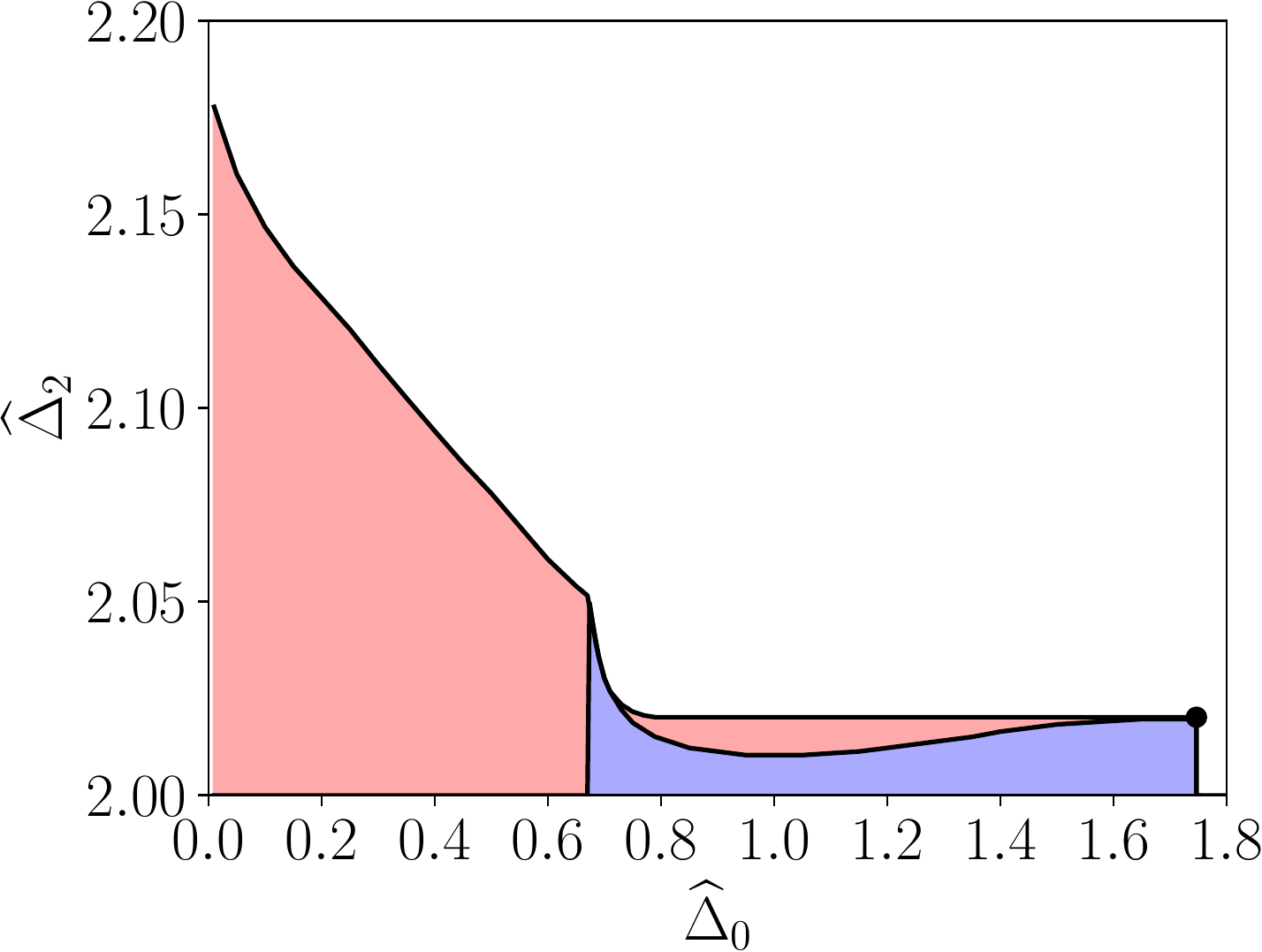}}\\
	\subfloat[$a_{\phi^2} = -0.1$, $\widehat{O}_1^2$ allowed]{\includegraphics[width=0.45\textwidth]{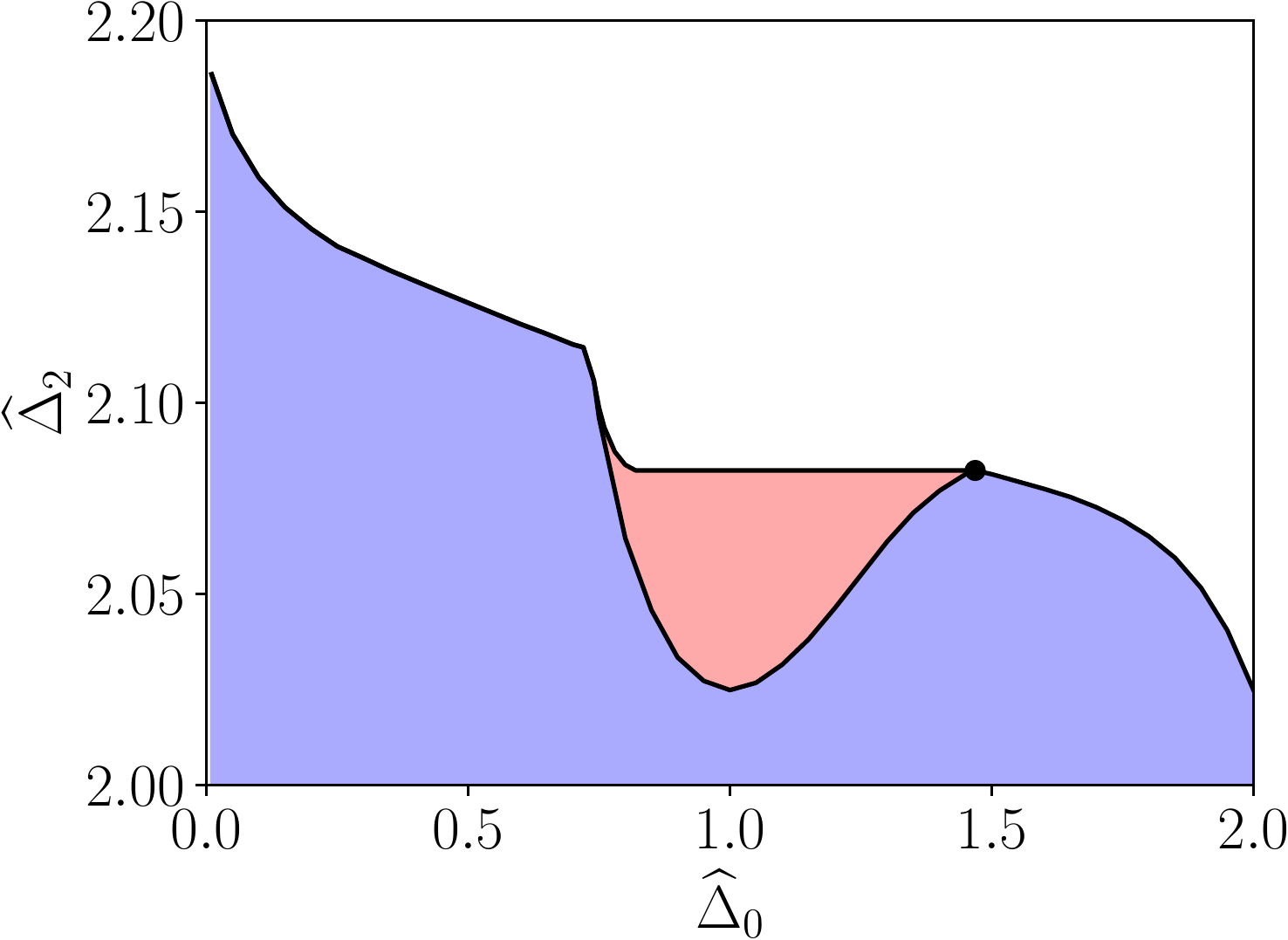}}
	\subfloat[$a_{\phi^2} = -0.4$, $\widehat{O}_1^2$ allowed]{\includegraphics[width=0.45\textwidth]{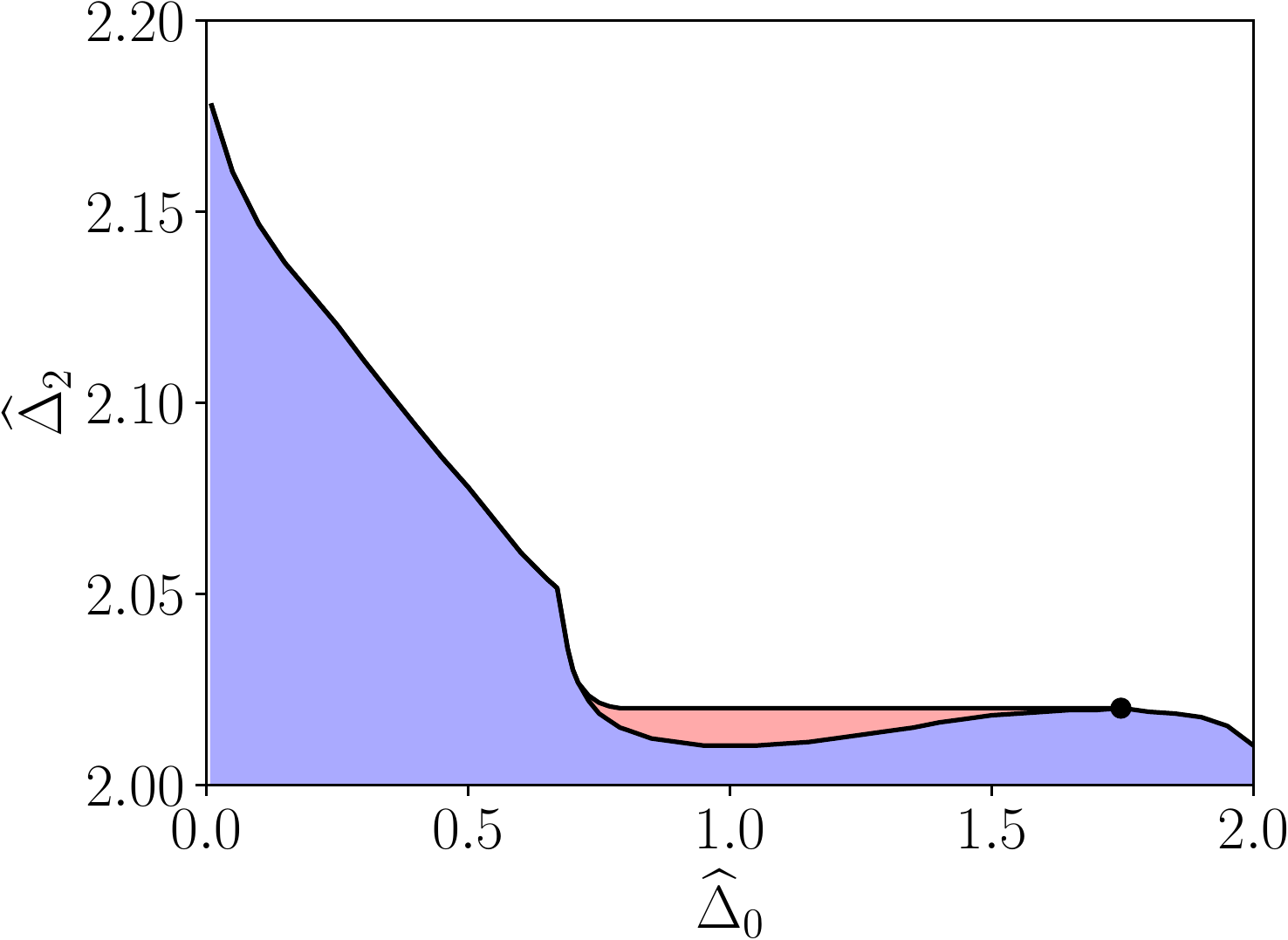}}
	\caption{Allowed points in $(\widehat{\Delta}_0, \widehat{\Delta}_2)$ space for selected values of $a_{\phi^2}$. In the blue region, $\widehat{\Delta}_0$ is the dimension of the unique relevant scalar while $\widehat{\Delta}_2$ is the spin 2 gap. In the pink region, they are both gaps. The rightmost corner (indicated with a dot) is where we conjecture that a minimal boundary condition lives. For previous examples of corners which maximize two gaps, see \cite{brv13,blrv15,Baggio:2017mas,Dymarsky:2017xzb,Dymarsky:2018}. It would be interesting to find an interpretation of the second kink to the upper left. Note that the smallest blue region is obtained when all possibilities for a second relevant scalar, including the one that can only appear in $\widehat{O}_1 \times \widehat{O}_1$, are disallowed. If one only disallows a \textit{generic} second relevant scalar, the blue and pink regions become more similar.}
	\label{fixed-a}
\end{figure}

\subsubsection*{Region III}
We also discovered a second kink for lower values of $\hD_0$, marked with a III in figure \ref{fig:3dplot}. A good way to visualize this feature is to make plots along the lines of figure \ref{fixed-a} at fixed $a_{\phi^2}$.\footnote{Note that the pink regions of these plots correspond to slices in figure \ref{fig:3dplot}. The blue ones reflect different assumptions about the spectrum.} For each subfigure the rightmost kink corresponds to the minimal model boundary condition, and our interest now lies with the kink appearing for $\hD_0 < 1$.
\begin{figure}
        \centering
        \includegraphics[width=0.85\textwidth]{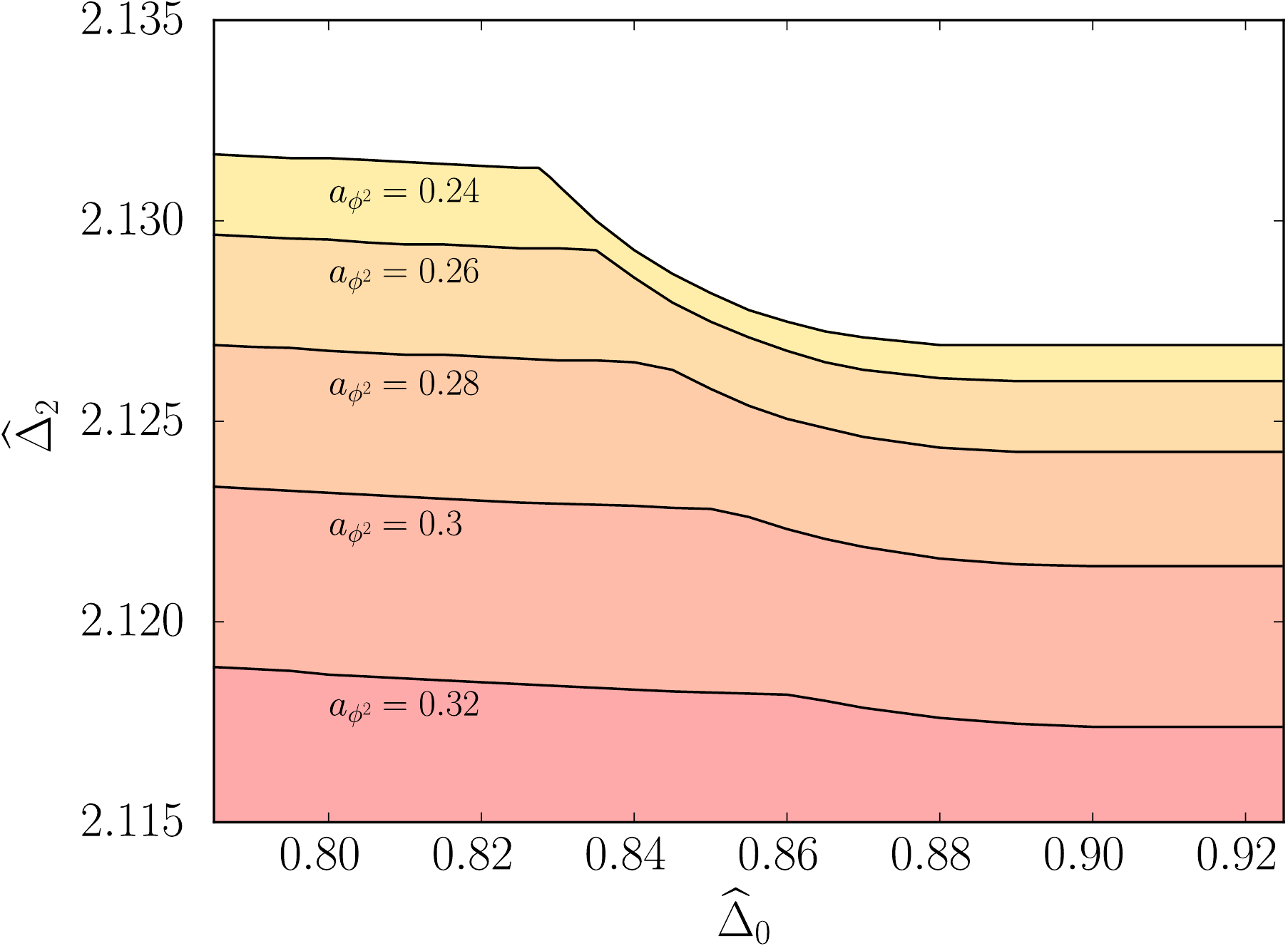}
        \caption{Allowed regions analogous to figure~\ref{fixed-a} which come from following the upper left kinks there to five larger values of $a_{\phi^2}$. For the nearly smoothed out boundary at $a_{\phi^2} = 0.32$, the two $\hD_2$ values of interest have almost converged. Notice that the sharp cliff corresponding to the minimal model boundary condition lies further to the right, outside the range of this plot.}
        \label{kink-series}
\end{figure}
The physical interpretation of this kink is an enigma. We have tried to follow it to the endpoints of the $a_{\phi^2}$ interval. For $a_{\phi^2} \to - 1/2$ it appears that $\hD_0 \to 0.66$, so approximately 2/3, and $\hD_2 \to 2$, but we do not have a natural CFT to associate to these dimensions. The fate of the kink upon increasing $a_{\phi^2}$ is shown in figure \ref{kink-series}. (The data for this plot was not shown in figure \ref{fig:3dplot} to avoid clutter.) We see that the kink ends up sinking under the minimal model boundary condition, but without merging with it; this can naturally occur because of the aforementioned convexity property of the bounds. Unfortunately this implies that we cannot track the fate of the kink for $a_{\phi^2} \to 1/2$.

We present an analysis of the spectrum at the kink below.

\subsection{Extremal spectra}
In the previous subsection we provided evidence that minimal boundary conditions maximize both the scalar and spin 2 gap. Theories that saturate bootstrap bounds are advantageous because they can be studied with the extremal functional method \cite{Poland:2011,ElShowk:2013}. Extremizing an OPE coefficient (while redundant at the precise boundary) can also lead to more stable results. We have carried this out by imposing gaps that correspond to kinks in $(a_{\phi^2}, \widehat{\Delta}_0, \widehat{\Delta}_2)$ space and maximizing $\hat{\lambda}^2_{12\widehat{\tau}}$ (where $\widehat{\tau}$ is the operator of dimension $\widehat{\Delta}_2$). The maximization output encodes the scaling dimensions and OPE coefficients of exchanged operators as described in \cite{Simmonsduffin:2017} whose script we also use.

To deal with the behaviour in region II, we have chosen to maximize $\hD_0$ first and $\hD_2$ second. This means that for $a_{\phi^2} > 0.36$ (where the selected point is actually unlikely to correspond to a physical boundary condition), we follow the upper branch in the right plot of figure~\ref{corner}. It turns out that the spectra in between the upper and lower branches only differ significantly in the dimension of their lightest scalar.

\begin{figure}
	\centering
	\subfloat[$l = 0$]{\includegraphics[width=0.45\textwidth]{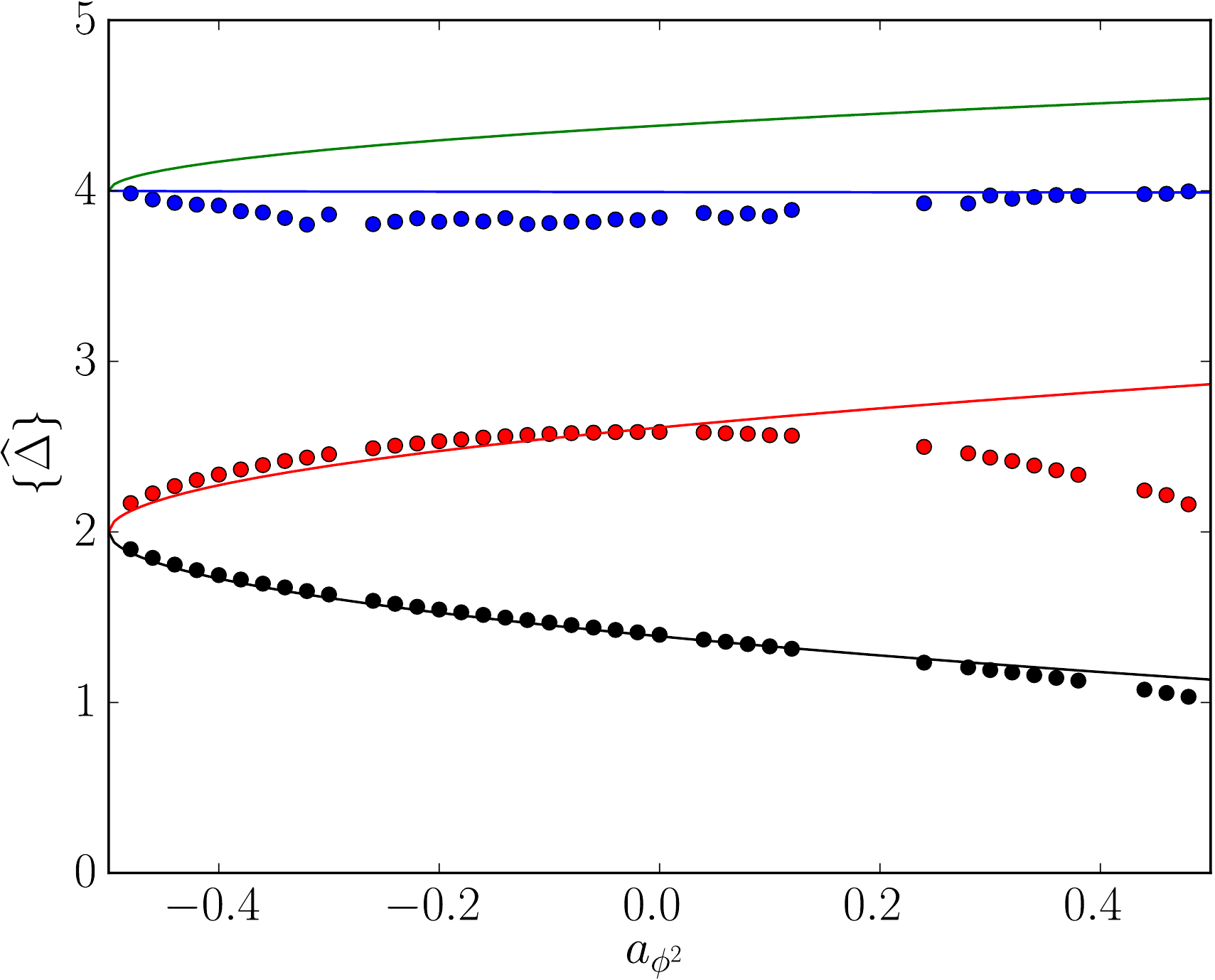}}
	\subfloat[$l = 2$]{\includegraphics[width=0.45\textwidth]{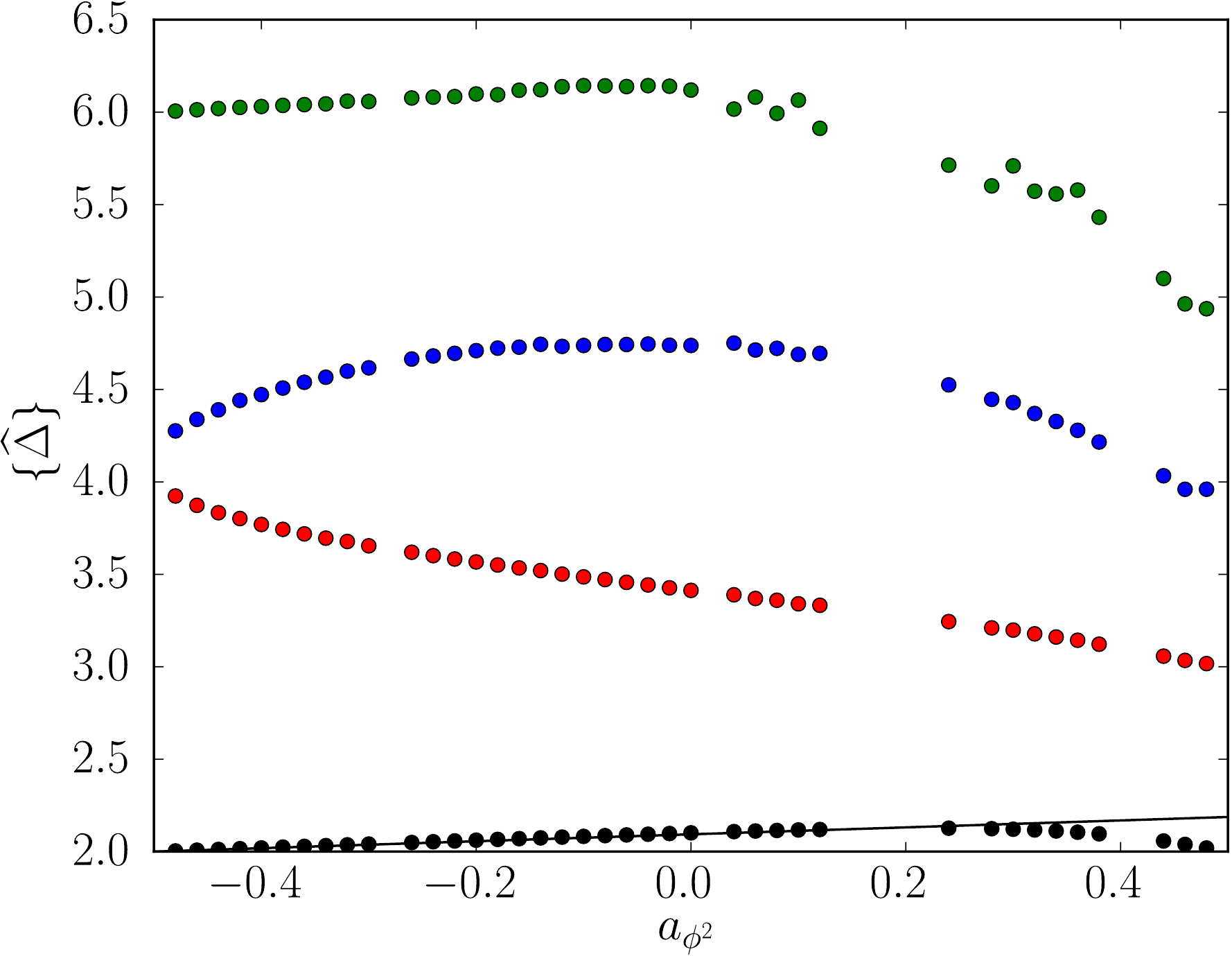}}
	\caption{Extremal spectra along the upper branch of figure~\ref{corner} for unprotected operators of spins 0 and 2. Alongside the data points we have also added perturbative results around the Dirichlet end (as a solid line of the same color) where available.}
	\label{spin02spectrum}
\end{figure}

\subsubsection*{The spectrum in regions I and II}
In figure \ref{spin02spectrum} we show operator dimensions for the first two spins as a function of $a_{\phi^2}$. Apart from those in the plot, we also assume the existence of spin 0 and spin 2 operators at the protected scaling dimensions of $3, 5, 7, \dots$ but only in the Bose symmetric OPEs.  The dimension of the lowest operator in either channel is (of course) exactly the maximal gap $\hD_0$ or $\hD_2$, so the new information in these plots lies in the higher-dimensional operators.

For the lowest pairs of operators the numerical spectrum in figure \ref{spin02spectrum} matches well against the perturbative results. Note also that the spectrum starts and ends at integer values, in agreement with the expectations for the Dirichlet and Neumann points. On the other hand, we show in appendix \ref{app:moredata} that there are two scalars of dimension $4 + O(1 / m)$ but the functional only sees one of them. This likely happens because the other has a parametrically smaller OPE coefficient. We remind the reader that the extremal functional method is not an exact method, so small such discrepancies are not unexpected.

Another feature of figure \ref{spin02spectrum} is the smooth behavior of the spectrum as a function of $a_{\phi^2}$. (The same smoothness will be visible below when we discuss the OPE coefficients for some of the protected operators.) We take this as another indication that the $m=3$ minimal boundary condition is just the Neumann point. Indeed, suppose this boundary condition would correspond to a non-trivial fixed point with $a_{\phi^2}$ strictly smaller that $1/2$. Then it is highly likely that we would have seen some rearrangement in the extremal spectrum at this point. More precisely, for $m=3$ the $\Phi_{1,3}$ operator gets a null descendant at level 2 (because it is then equal to $\Phi_{2,1}$) which we can schematically denote as $L_{-2} \Phi_{1,3}$. This operator has spin 2 and dimension 4 for large $m$. In figure \ref{spin02spectrum} we however see no sign of such a rearrangement: both operators that have these quantum numbers at $a_{\phi^2} = -1/2$ continue to exist smoothly all the way down to the Neumann point with $a_{\phi^2} = +1/2$.

The extremal function method also gives us access to protected operators which are only in $\widehat{O}_1 \times \widehat{O}_1$ and $\widehat{O}_2 \times \widehat{O}_2$. The coefficients of the scalars in these OPEs are especially interesting because they allow us to check the displacement Ward identity, along with its higher-spin counterpart, \textit{a posteriori}.

\subsection*{The first two Ward identities in regions I and II}
To make sure that the spectra in figure~\ref{spin02spectrum} correspond to local boundary conditions, we need to be able to see artifacts of the bulk currents. In particular, one of the dimension $3$ scalars must be $\Disp_2^{(0)} \equiv \Disp$ and one of the dimension $5$ scalars must be $\Disp_4^{(0)}$. In all the extremal solutions we obtained (by maximizing $\hat{\lambda}^2_{12\widehat{\tau}}$) the dimension $3$ and $5$ scalars turn out to be unique, so it is straightforward to read off the numerical predictions for $\hat{\lambda}_{11\Disp}$, $\hat{\lambda}_{22\Disp}$, $\hat{\lambda}_{11\Disp_4^{(0)}}$ and $\hat{\lambda}_{22\Disp_4^{(0)}}$. Recalling our conventions in appendix~\ref{app:conventions}, these are given by
\begin{align}
        \hat{\lambda}_{ii\Disp} \equiv \frac{\hat{f}_{ii\Disp}}{\sqrt{C_{\Disp}}}~,\quad \hat{\lambda}_{ii\Disp_4^{(0)}} \equiv \frac{\hat{f}_{ii\Disp_4^{(0)}}}{\sqrt{C_{\Disp_4^{(0)}}}}
\end{align}
for as yet unknown central charges.

\begin{figure}
        \centering
        \subfloat[$\ell = 2$]{\includegraphics[width=0.45\textwidth]{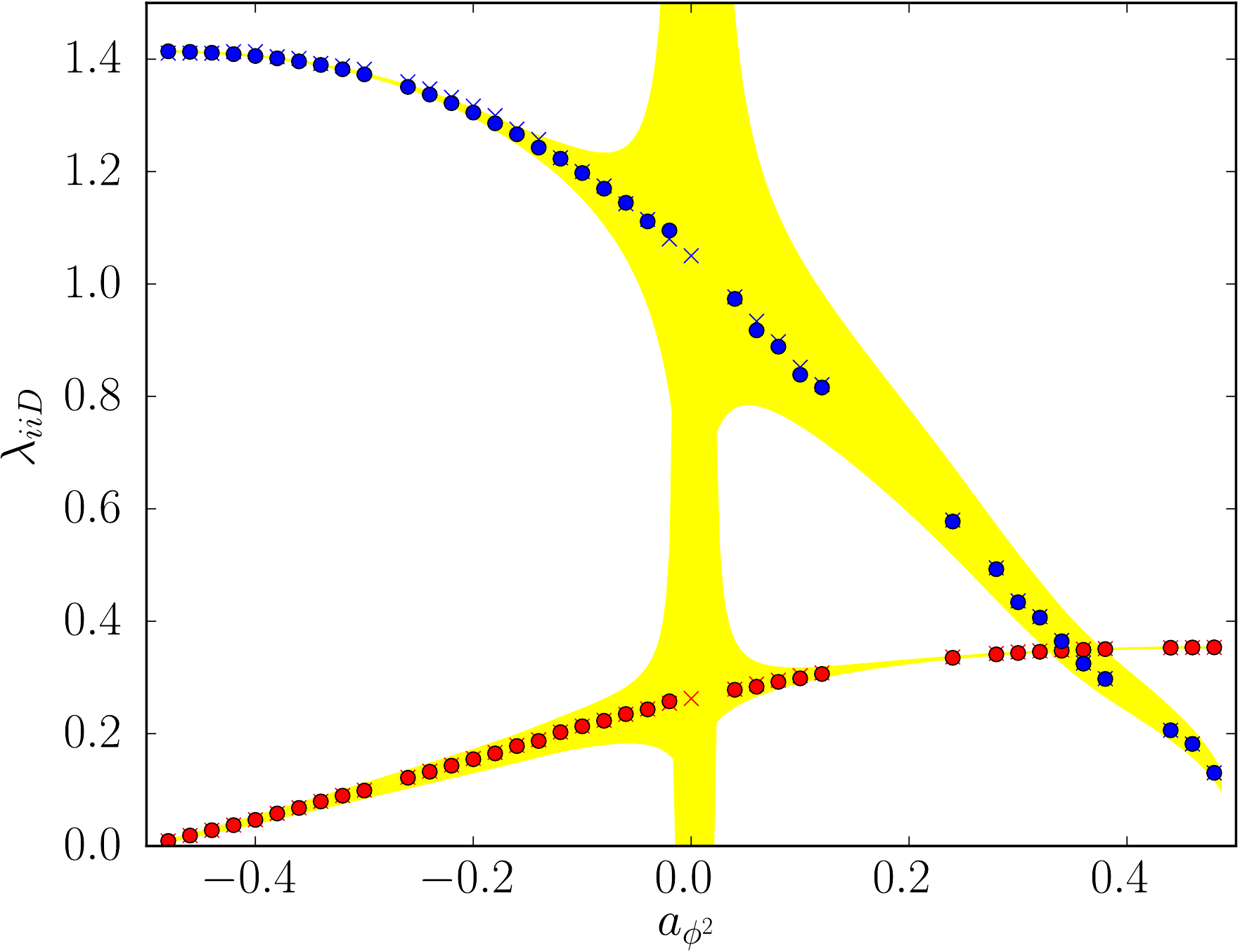}}
        \subfloat[$\ell = 4$]{\includegraphics[width=0.45\textwidth]{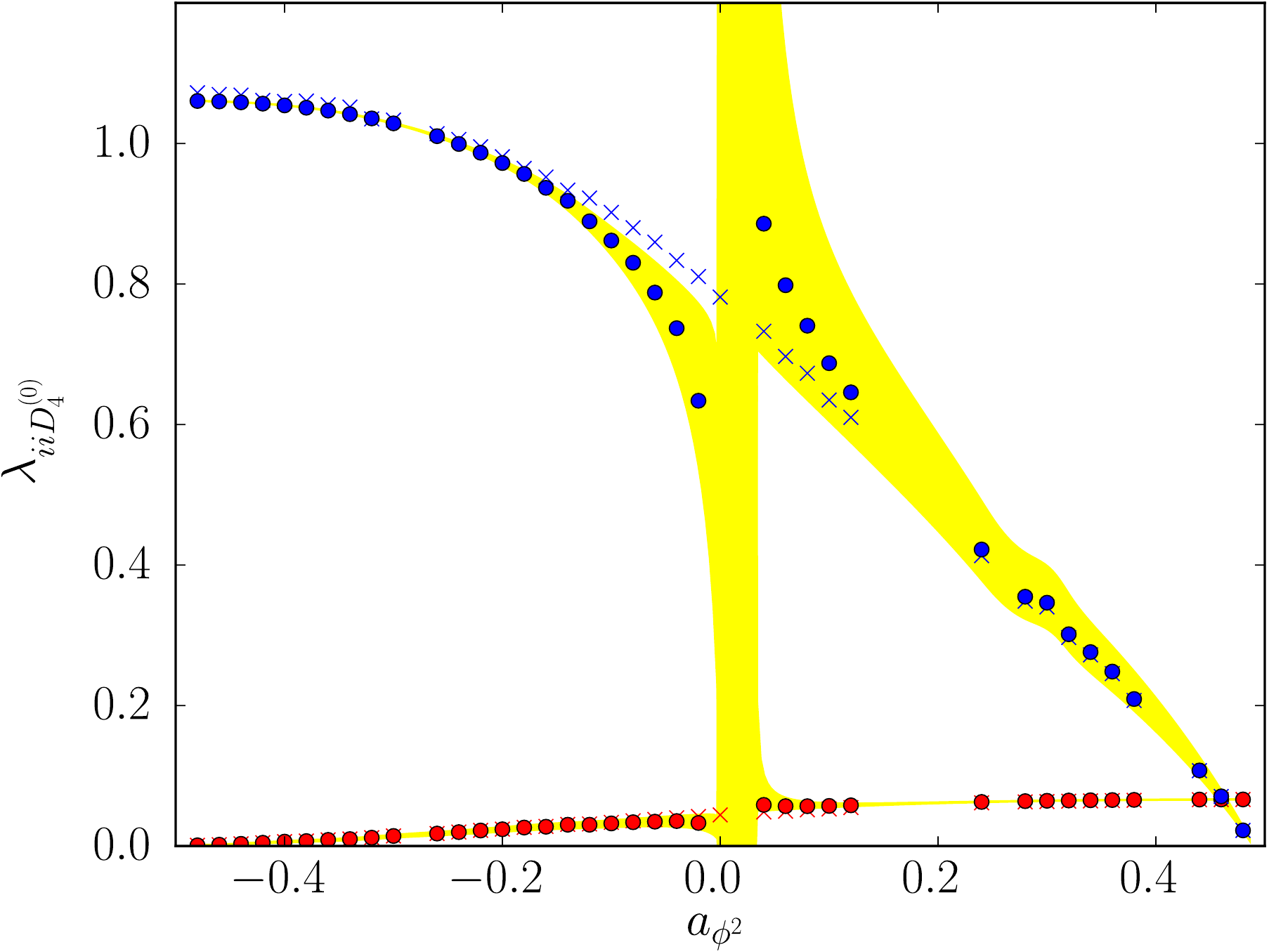}}
        \caption{Plots demonstrating the extent to which the Ward identities hold at the kinks. Red and blue crosses are the numerically determined $\lambda_{11\Disp_\ell^{(0)}}$ and $\lambda_{22\Disp_\ell^{(0)}}$ respectively. The red and blue circles are obtained by computing a displacement central charge using their ratio and plugging this back into \eqref{WardD20} and \eqref{WardD40}. The yellow envelope shows how much the circles can vary if the OPE coefficients are allowed to fluctuate within $5\%$ of their extracted values. As such, large disagreements should not be concerning if they occur near $a_{\phi^2} = 0$. Since the extremal functional method does not yield rigorous errors, $5\%$ was chosen arbitrarily.}
        \label{wardid}
\end{figure}

The consequences of locality for the displacement operator $\Disp$ were derived in \cite{Behan:2020nsf} and reviewed in subsection \ref{wardids}. Due to the Ward identity, the two OPE coefficients must obey the relation in eq.~\eqref{WardD20} which overdetermines the value of $C_\Disp$. 
In \cite{Behan:2020nsf}, this relation was checked at a single point. Since our current task is to check it over the full range of $a_{\phi^2}$, we must contend with the fact that there are several options for measuring the violation of the Ward identity. As an example, one could choose $C_\Disp$ by demanding that \eqref{WardD20} reproduce the numerically obtained $\hat{\lambda}_{11\Disp} \sqrt{C_\Disp}$ and then check whether the same is true for $\hat{\lambda}_{22\Disp} \sqrt{C_\Disp}$. Following this approach would lead to a quadratic equation. For simplicity, we will extract $C_\Disp$ by considering the ratio $r_\Disp \equiv \hat{f}_{11\Disp} / \hat{f}_{22\Disp} = \hat{\lambda}_{11\Disp} / \hat{\lambda}_{22\Disp}$ and therefore only encounter a linear equation. From the solution (for $d = 3$) we find that:
\begin{equation}\label{solCD}
        C_\Disp = \frac{4a_{\phi^2}}{S_d^2} \frac{4b_1^2 r_\Disp + b_2^2}{4b_1^2 r_\Disp - b_2^2}\,.
\end{equation}
Having thus estimated $C_\Disp$ from the \emph{ratio} of the OPE numerical coefficients, it is straightforward to check how well the \emph{individual} OPE coefficients actually agree with \eqref{WardD20}. This way of verifying the Ward identity leads to the comparison in the left plot of figure \ref{wardid}.

An analogous situation applies to the higher displacement operators. For $\Disp_4^{(0)}$, the relations were  written in eq.~\eqref{WardD40}. 
We can again use their ratio to solve for $C_{\Disp_4^{(0)}}$. Defining $r_{\Disp_4^{(0)}} \equiv \hat{f}_{11\Disp_4^{(0)}} / \hat{f}_{22\Disp_4^{(0)}} = \hat{\lambda}_{11\Disp_4^{(0)}} / \hat{\lambda}_{22\Disp_4^{(0)}}$,
\begin{equation}\label{solCD4}
        C_{\Disp_4^{(0)}} = \frac{36a_{\phi^2}}{S_d^2} \frac{16b_1^2 r_{\Disp_4^{(0)}} + b_2^2}{16b_1^2 r_{\Disp_4^{(0)}} - b_2^2}
\end{equation}
is the central charge for a given dataset. By plugging this back into \eqref{WardD40}, we can again check the Ward identity by comparing the result to individual OPE coefficients in the numerics. This is done in the right plot of figure~\ref{wardid}.

\begin{figure}
        \centering
        \includegraphics[width=0.85\textwidth]{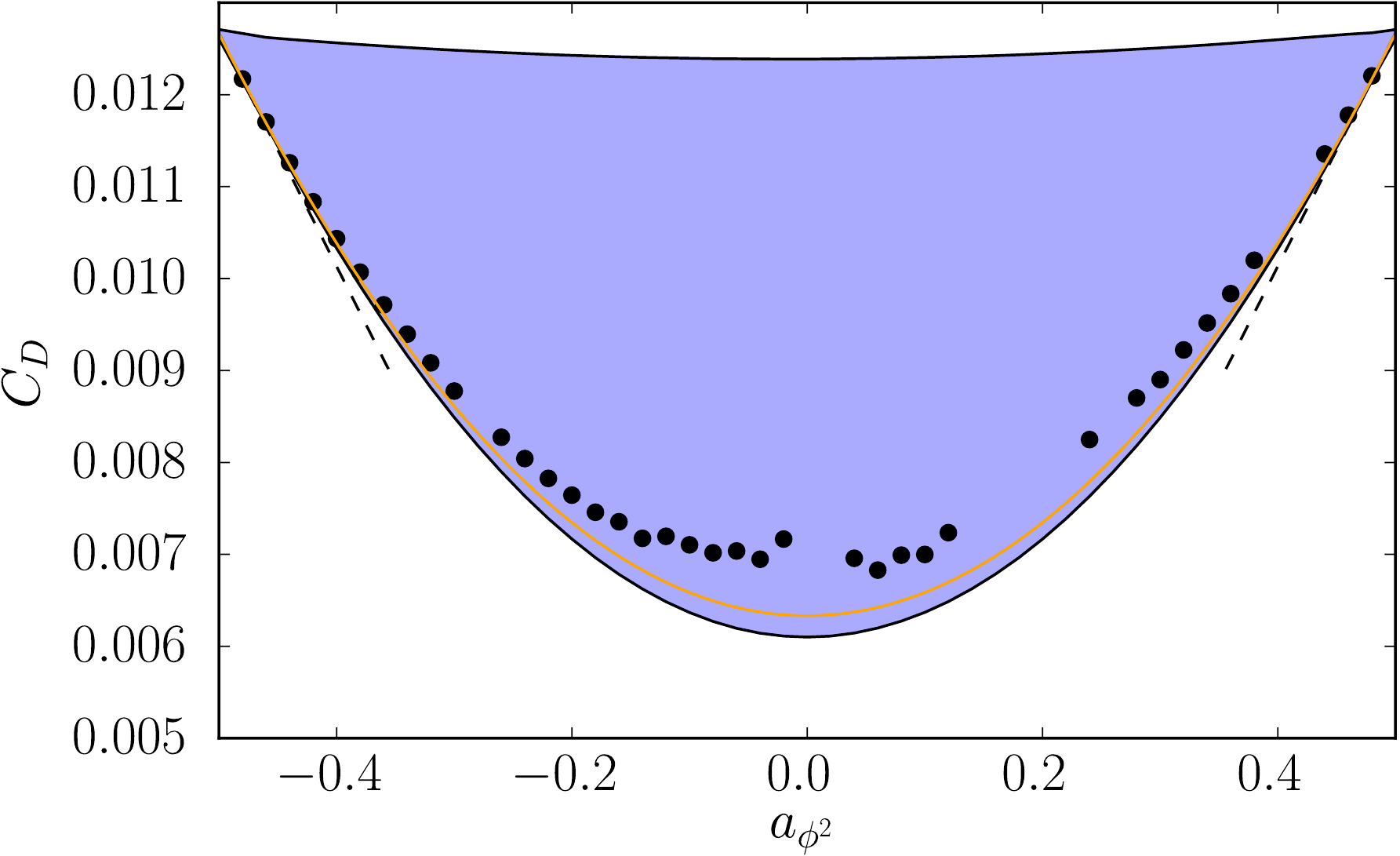}
        \caption{The allowed region for the displacement central charge with minimal model b.c. results (extracted with the extremal functional method) given by black dots. The dotted lines on either side are the first-order predictions of conformal perturbation theory while the orange line in between is the $C_\Disp$ versus $a_{\phi^2}$ curve realized by \eqref{coupled-gff}.}
        \label{crescent}
\end{figure}
An important property of equations \eqref{solCD} and \eqref{solCD4} is the potential divergence when $r_\Disp = \frac{b_2^2}{4b_1^2}$ and when $r_{\Disp_4^{(0)}} = \frac{b_2^2}{16b_1^2}$, respectively. In actual theories these values must occur precisely at $a_{\phi^2} = 0$ to ensure that $C_\Disp > 0$. This will however never quite happen in a numerical spectrum, and plots of \eqref{WardD20} and \eqref{WardD40} with sufficient resolution will necessarily encounter a pole. In other words, small numerical violations of the Ward identities near $a_{\phi^2} = 0$ can lead to huge discrepancies between the predicted and numerical OPE coefficients. That is why the error window also diverges in figure \ref{wardid}. The right plot of figure~\ref{wardid} explictly shows this pole in the data. It is fairly wide because our numerical estimates of the $\Disp_4^{(0)}$ OPE coefficients at $a_{\phi^2} = 0$ have a ratio of $0.05672$ --- about $10\%$ away from the ideal value of $1/16$. Looking at $\Disp$, on the other hand, the same ratio at $a_{\phi^2} = 0$ is $0.24988$. Since this is only $0.05\%$ away from $1/4$, the pole on the left plot is too narrow to see.

Outside of the window near $a_{\phi^2} = 0$ our results show an excellent match with the Ward identities, providing another confirmation that these solutions to the crossing equations correspond to physical boundary conditions.

Finally, it is interesting to see how the numerical results for \eqref{solCD} compare to the minimum and maximum allowed values of $C_\Disp$ for a given $a_{\phi^2}$. This is shown in figure \ref{crescent} where the bounds come from adding explicit blocks with $\hat{\lambda}_{ii\Disp}$ coefficients and scanning over $C_\Disp$ to see when crossing can and cannot be solved. Since there is no gap between the identity and the other boundary scalars, it is not obvious that such bounds need to exist in the first place. Any parameter changing $\hat{\lambda}_{11\Disp}$ and $\hat{\lambda}_{22\Disp}$ by an overall rescaling, for example, would have been completely unconstrained. We therefore attribute figure \ref{crescent} to the presence of multiple correlators in our setup and the fact that $C_\Disp$ enters the Ward identity in a non-linear way.

In contrast to the analogous 4d/3d plot in \cite{Behan:2020nsf}, the upper bound on $C_\Disp$ here is symmetric. We also see that the lower bound is asymptotic to the perturbative results around the Neumann and Dirichlet ends which are given by \eqref{aphi2CD} and \eqref{DminimalcptFirstCdvsa} respectively. There is an exactly solvable theory, interpolating between Dirichlet and Neumann, which comes remarkably close to the lower bound at intermediate values of $a_{\phi^2}$. This is given by coupling the Dirichlet bulk field to a GFF $\widehat{\chi}$ having dimension $\hD_{\widehat{\chi}} = d/2 - 1$ via
\begin{equation}
\delta S_\partial = g \int_{y=0} \,\mathrm{d}^{d-1}\vec{x}  \,  \partial_y \phi \, \widehat{\chi}. \label{coupled-gff}
\end{equation}
Being a coupling between two GFFs with shadow dimensions, $g$ is exactly marginal \cite{w02}.\footnote{In the context of BCFT this was considered in \cite{DiPietro:2020fya,Herzog:2021spv}. The resulting line of (non-local) CFTs admits a dual description in terms of a linear deformation of Neumann with coupling $g'=1/g$ \cite{w02}.}
 Since the theory is gaussian, the one-point function of $\phi^2$ and the two-point function of $\Disp$ are then straightforwardly computed, and for any $g$ they are related by
\begin{align}\label{aphi2CDGFF}
C_{\Disp}(g)&=\frac{\Gamma \left(\frac{d}{2}\right)^2}{4\pi^{d}} \left(1+a_{\phi^2}^2(g) 4^{d-2}\right)~.
\end{align}
It is pleasing that this solution appears to hug the lower bound on $C_\Disp$ both for $d = 3$ (as in figure \ref{crescent}) and for $d = 4$ \cite{Behan:2020nsf}. We expect this agreement to become arbitrarily good upon increasing the number of derivatives. As a check, solutions along the lower bound feature a flux operator of the type found in \eqref{coupled-gff}.

\begin{figure}
        \centering
        \subfloat[$l = 0$]{\includegraphics[width=0.45\textwidth]{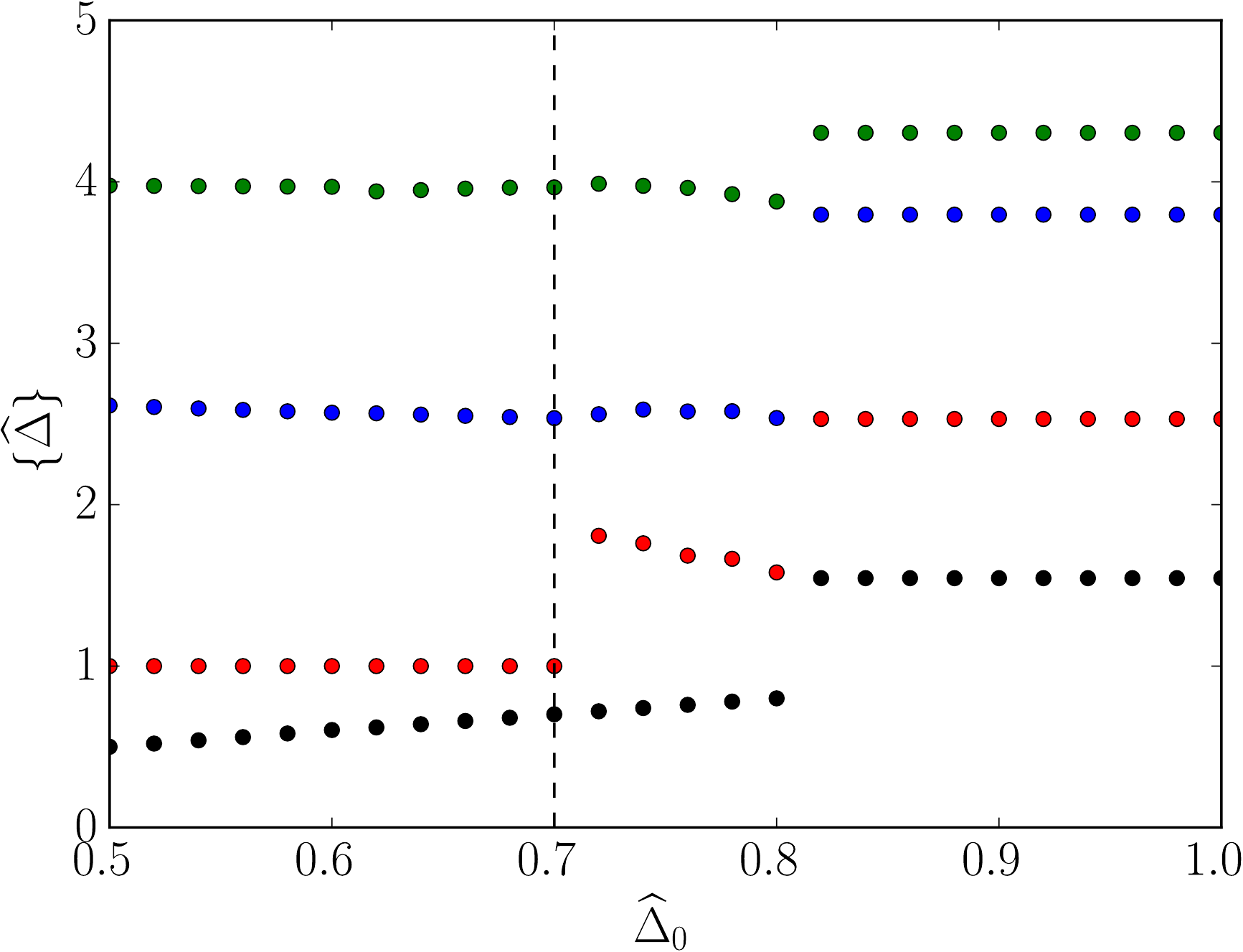}}
        \subfloat[$l = 2$]{\includegraphics[width=0.45\textwidth]{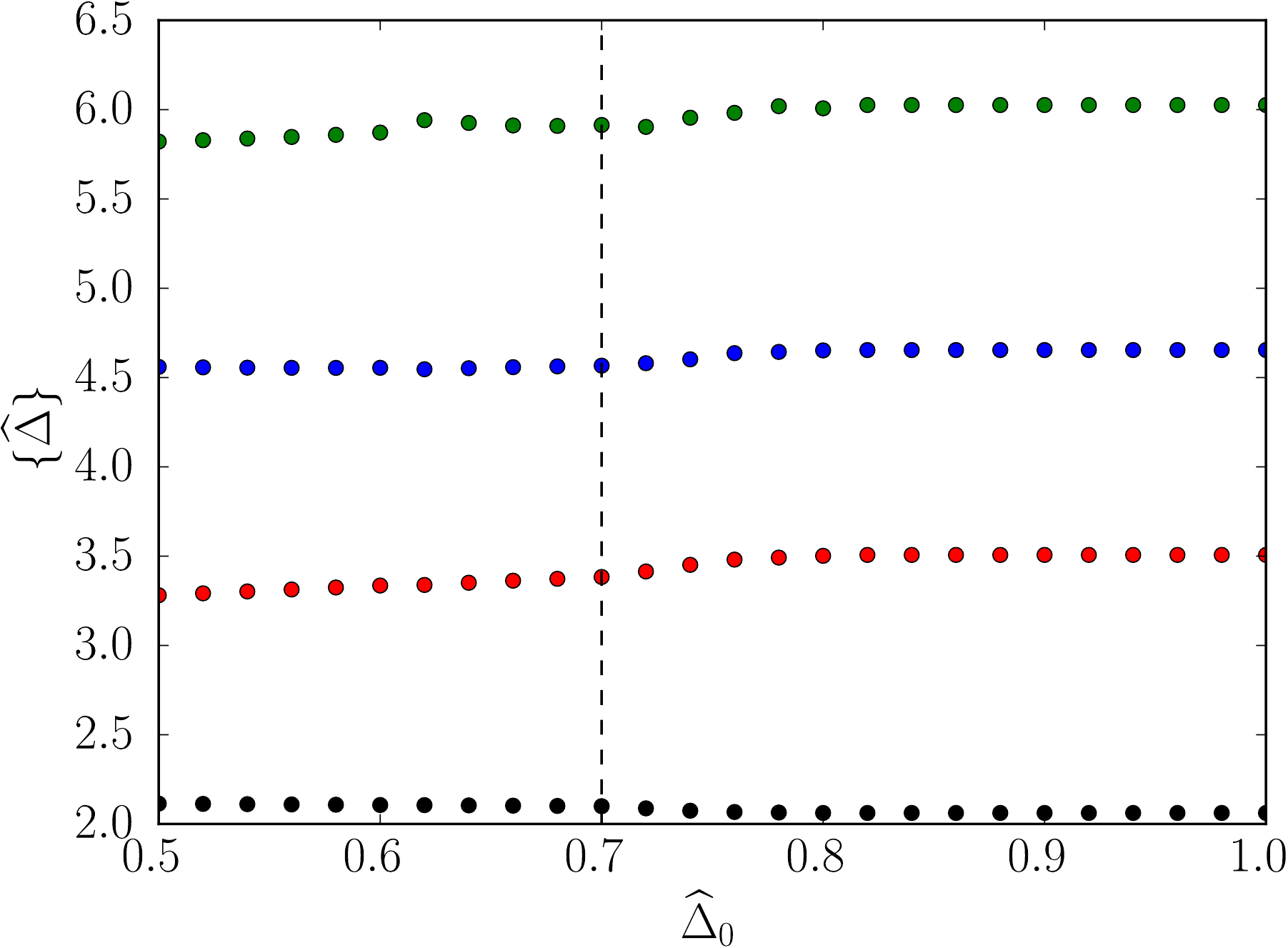}}
        \caption{Spectra in the vicinity of the exotic kink for $a_{\phi^2} = -0.2$. For each value of the imposed scalar gap $\hD_0$, we plot the dimensions of the first four operators of spins 0 and 2 for the solutions which maximize the gap $\hD_2$. The location of the kink itself is indicated with a vertical dotted line.}
        \label{spin02other}
\end{figure}

\subsubsection*{The spectrum in region III}
We also used the extremal functional method for the exotic kinks in region III. This time, extracting the spectrum in the right place requires more care. In figure \ref{spin02other} we plot the spectrum along the boundary of the allowed region in $(\hD_0, \hD_2)$ space for the fixed value $a_{\phi^2} = -0.2$. The exotic kink here happens to be very close to $\hD_0 = 0.7$, although the drop in the dimension of the first spin 2 operator is not visible at the scale of this plot.

For $\hD_0 > 0.8$ we see that the spectrum remains constant and in fact becomes equal to that of the minimal model kink. In other words, for these values there is no operator exactly at the value of the imposed gap $\hD_0$ and we learn nothing new.

For $\hD_0 \approx 0.8$ we see a rearrangement of the spectrum and the extremal solution does require an operator at the imposed gap. Moving further to the left towards the other kink at $\hD_0 \approx 0.7$ the situation is more interesting. This time it is the second lightest scalar whose OPE coefficient decreases. But as soon as it goes to zero, we find that $\widehat{O}_1^2$ of dimension $1$ steps in to replace the missing operator.

It is curious that for all of the operators we have looked at other than the first two scalars, the dimensions never change by more than $10\%$ over the whole range of figure \ref{spin02other}. This continues to be the case for other values of $a_{\phi^2}$ -- if the numerics find an irrelevant operator at the minimal model kink, they will find one having almost the same dimension at the other kink. This in particular implies that both exhibit a weakly irrelevant operator as we take $a_{\phi^2} \to -1/2$. The exotic kink does not have a weakly relevant operator in this limit due to the decoupling of the red / black line. However, the dimension of its most strongly relevant operator appears to be approachig $2/3$.

Although not shown, we have also investigated the spectrum near the `sinking' of the kink as in figure \ref{kink-series}. For these values of $a_{\phi^2}$, we in particular traced the lightest scalar in the minimal boundary condition, so the scalar corresponding to the the red / black line in figure \ref{spin02other}. This scalar again tends upwards as we move to the left, but it eventually decouples well before dimension $2$. From figure \ref{kink-series} it is immediate that the lightest spin 2 operators have the same dimension when the exotic kink disappears into the convex space spanned by the minimal model kink. The remainder of the two spectra are however not identical: indeed, the lightest exchanged scalar at the exotic kink still has a significant OPE coefficient when the sinking takes place. Altogether it appears that the exotic kink is moving into the interior of the allowed region. It is possible that this happens at $a_{\phi^2} \approx 0.36$, which played a special role in our previous analysis, but we leave a detailed investigation for future work.

\section{Outlook}
\label{sec:conclusions}

The world of conformal boundaries and defects for free theories is remarkably rich. In this paper we focused on boundary conditions for a three-dimensional free scalar and found that there exist `minimal model boundary conditions' that we can analyze both perturbatively and numerically. A second kink in our numerical bounds remains unexplained.\footnote{We stress that our results are valid also if the scalar is compact, as would be the case for the Goldstone boson of a broken $U(1)$ symmmetry. For example, our minimal model boundary conditions are obtained by a coupling to the operator $\partial_y \phi$ which always exists.}

There are many evident directions for future research. Remaining within the realm of a free scalar bulk theory, we can continue numerical bootstrap studies for conformal defects of varying spacetime dimension $d$ and co-dimension $q$.

First let us consider taking the co-dimension $q$ to be non-integral. In that case the analysis of \cite{Behan:2017dwr, Behan:2017emf} indicates that for $q$ close to one our setup for $m = 3$ will flow to the long-range Ising model, which should be smoothly connected to the Neumann boundary condition as $q \to 1$. For $m > 3$ this naturally raises the question whether there are similar `long-range minimal models' that we can obtain by coupling the minimal model $\Phi_{1,2}$ operator to the bulk scalar for different values of $q$.

To expand on this idea, let us recall the Landau-Ginzburg description (see Section 7.4.7 of~\cite{DiFrancesco:1997nk} for a review) of the $m$'th minimal models as the fixed point of
\begin{equation}
	\frac{1}{2}(\partial \Phi)^2 + \Phi^{2m - 2}.
\end{equation}
Following the ideas of \cite{Behan:2017dwr,Behan:2017emf} we may want to couple $\Phi$ to a generalized free field, say $\partial_y \phi$ for a bulk scalar $\phi$ which extends to a $q$ co-dimensional space. In the infrared the fundamental scalar $\Phi$ however becomes the $\Phi_{2,2}$ operator, whereas we considered a coupling to the $\Phi_{1,2} \equiv \Phi_{m-1,m-1}$ operator. This latter operator should originate from $\Phi^{m-2}$ in the Landau-Ginzburg description and therefore we are rather looking at the fixed point of:
\begin{equation}
	\frac{1}{2}(\partial\Phi)^2 + \Phi^{2m - 2} + \Phi^{m-2} \partial_y \phi.
\end{equation}
It would be interesting to see whether such a theory can also flow to a non-trivial infrared fixed point for non-integer co-dimensions. Similarly we would like to understand the fate in the infrared of a more conventional long-range setup where the latter coupling gets replaced by $\Phi\, \partial_y \phi$ for any $m$.

Finally we would also like to try varying both the boundary and bulk dimensions in order to further understand the exotic kink in region III, the hope being that this kink will move into a perturbatively understandable region. 

Natural other directions are obtained by including more scalars or fermions in the bulk theory (see \cite{Giombi:2019enr,Prochazka:2019fah} for a study of $O(N)$ invariant boundary conditions for multiple bulk free massless scalars.). One may for example expect a rich variety of non-trivial supersymmetric boundary conditions. Alternatively, one could go in the direction of breaking even more symmetry for example by considering intersecting boundaries (see~\cite{Antunes:2021qpy} for a recent study along this direction). It would be also interesting to consider the Weyl rescaling of our minimal model boundary conditions to AdS, along the lines of \cite{Paulos:2016fap, Carmi:2018qzm, Herzog:2019bom, Giombi:2020rmc, Giombi:2021uae, Giombi:2021cnr}. The addition of a bulk mass in this setup would be another way to realize the coupling of the minimal models to a more general GFF.

\section*{Acknowledgements }

We would like to thank Davide Gaiotto for suggesting the coupling \eqref{eq:mmlag} to minimal models and Adam Nahum for discussions. EL and BvR are supported by the Simons Foundation grant $\#$488659 (Simons Collaboration on the non-perturbative bootstrap). LD thanks the participants of the workshop ``Topological properties of gauge theories and their applications to high-energy and condensed-matter physics'' at GGI, Florence, where this work was presented for interesting comments.
LD is partially supported by INFN Iniziativa Specifica ST\&FI. LD also acknowledges support by the program ``Rita Levi Montalcini'' for young researchers. CB has received funding from the European Research Council (ERC) under the European Union's Horizon 2020 research and innovation programme (grant agreement $\#$787185). The numerical calculations were done on the University of Oxford Advanced Research Computing (ARC) facility \cite{Richards:2015}.

\appendix

\section{Conventions}
\label{app:conventions}

\subsection{bOPE}
Consider a scalar bulk operator $\mathcal{O}$, not necessarily free. 
The bOPE of $\mathcal{O}$ is completely determined by $SO(d,1)$ symmetry, up to a certain collection of CFT data \cite{McAvity:1993ue,McAvity:1995zd}. Up to boundary descendants this reads
\begin{align}\label{bOPEphi2}
	\mathcal{O}(\vec{x},y)=\sum_{\widehat{\mathcal{O}}} \frac{b_{\mathcal{O}}{}^{\widehat{\mathcal{O}}}}{y^{\Delta_{\mathcal{O}} -\widehat{\Delta}_{\widehat{\mathcal{O}}}}}\widehat{\mathcal{O}}(\vec{x})+\dots~.
\end{align}
The boundary operators $\mathcal{O}$ are scalar primaries with two-point correlation functions given by
\begin{align}\label{Btwopt}
	\langle \widehat{\mathcal{O}}(\vec{x}) \widehat{\mathcal{O}}'(0)\rangle =\frac{\widehat{C}_{\widehat{\mathcal{O}}\widehat{\mathcal{O}}'}}{|\vec{x}|^{2\widehat{\Delta}_{\widehat{\mathcal{O}}}}}~, \quad \widehat{C}_{\widehat{\mathcal{O}}\widehat{\mathcal{O}}'}=\delta_{\widehat{\mathcal{O}}}^{\widehat{\mathcal{O}}'}\widehat{C}_{\widehat{\mathcal{O}}\widehat{\mathcal{O}}}~,
\end{align}
The contributions from the boundary descendants are determined by the bulk-boundary correlators 
\begin{align}\label{BD2pt}
	\langle \mathcal{O}(\vec{x},y)\widehat{\mathcal{O}}(0)\rangle =\frac{b_{\mathcal{O}\widehat{\mathcal{O}}}}{y^{\Delta_{\mathcal{O}}-\widehat{\Delta}_{\widehat{\mathcal{O}}}}(|\vec{x}|^2+y^2)^{\widehat{\Delta}_{\widehat{\mathcal{O}}}}}~,
\end{align}
where
	 $b_{\mathcal{O}}{}^{\widehat{\mathcal{O}}}\widehat{C}_{\widehat{\mathcal{O}}\widehat{\mathcal{O}}}=b_{\mathcal{O}\widehat{\mathcal{O}}}$. 
We will take unit-normalized boundary two-point functions, except for the protected operators that can appear in the bOPE of the bulk conserved currents $J_\ell$. Such operators, collectively denoted by $\widehat{J}_\ell^{(l)}$ (with $l=0,\dots \ell-1$) have their normalization fixed by the Ward identities \eqref{ward_currents}, and therefore the coefficients in their two-point functions are physical
\begin{align}
	\begin{split}\label{JhDtwopt}
		\langle \Disp(\vec{x})\Disp(0)\rangle &=  \frac{C_{\Disp}}{|\vec{x}|^{2d}}~,\\
		\langle \widehat{{J}}^{(l)}_\ell(\vec{x},z_1)\widehat{{J}}^{(l)}_\ell(0,z_2)\rangle & = {C}_{\widehat{J}^{(l)}_\ell} \frac{(z_1 \cdot I(\hat{x})\cdot z_2)^l}{|\vec{x}^2|^{d+\ell-2}}~.
	\end{split}
\end{align}

\subsection{Boundary OPE and physical OPE coefficients}
Consider the OPE between two boundary operators $\widehat{\mathcal{O}}_i$. 
Up to boundary descendant this reads
\begin{align}\label{boundaryOPEgen}
	\widehat{\mathcal{O}}_i(\vec{x})\widehat{\mathcal{O}}_j(0)\sim \sum_k\frac{\hat{f}_{ij}{}^{k}}{|\vec{x}|^{\widehat{\Delta}_i+\widehat{\Delta}_j-\widehat{\Delta}_k}}\widehat{\mathcal{O}}_k(0)+\dots
\end{align}
The boundary two-point functions are normalized as in \eqref{Btwopt}. We use the Zamolodchikov metric $\widehat{C}_{\widehat{\mathcal{O}}\widehat{\mathcal{O}}'}$ to raise and lower indices of $\hat{f}_{ij}{}^{k}$s as follows
\begin{align}
	\langle \widehat{\mathcal{O}}_i (\vec{x}_1)\widehat{\mathcal{O}}_j (\vec{x}_2)\widehat{\mathcal{O}}_m (\infty)\rangle=\frac{{}\hat{f}_{ij}{}^k \widehat{C}_{km}}{|x_{12}|^{\widehat{\Delta}_i+\widehat{\Delta}_j-\widehat{\Delta}_m}}\equiv\frac{\hat{f}_{ijm}}{|x_{12}|^{\widehat{\Delta}_i+\widehat{\Delta}_j-\widehat{\Delta}_m}}~.
\end{align}
The displacement operator, whose normalization is taken as in \eqref{JhDtwopt}, enters the generic boundary OPE \eqref{boundaryOPEgen} as
\begin{align}
	\widehat{\mathcal{O}}_i(x) \widehat{\mathcal{O}}_j(0) &\supset \,\frac{{\hat{f}_{ij}{}^{\Disp}}}{|x|^{\widehat{\Delta}_i+\widehat{\Delta}_j-d}}\Disp(0)+\dots,
\end{align}
and a generic boundary four-point function as
\begin{align}
	\begin{split}
		\langle \widehat{\mathcal{O}}_i(0) \widehat{\mathcal{O}}_j(x)\widehat{\mathcal{O}}_k(1) \widehat{\mathcal{O}}_m(\infty)\rangle &\supset		 {\hat{f}_{ij}{}^{\Disp}\hat{f}_{km}{}^{\Disp}}{}\langle\Disp(0)\Disp(\infty)\rangle(1+\dots)\\
		&={\hat{f}_{ij}{}^{\Disp}\hat{f}_{km\Disp}}\,\,g_{\Disp}^{\widehat{\Delta}_{ij},\widehat{\Delta}_{kl}}(u,v)\\
		&=\frac{{\hat{f}_{ij\Disp}\hat{f}_{km\Disp}}}{C_{\Disp}}\,\,g_{\Disp}^{\widehat{\Delta}_{ij},\widehat{\Delta}_{kl}}(u,v)~.
	\end{split}
\end{align}
In the equation above we introduced the conformal blocks in the standard normalization. It will be sometimes convenient choose $\Disp$ to be unit-normalized. In that case the physical boundary OPE coefficient is
\begin{align}
	\langle \widehat{\mathcal{O}}_i(0) \widehat{\mathcal{O}}_j(x)\widehat{\mathcal{O}}_k(1) \widehat{\mathcal{O}}_m(\infty)\rangle &\supset{{\hat{\lambda}_{ij\Disp}\hat{\lambda}_{km\Disp}}}{}\,\,g_{\Disp}^{\widehat{\Delta}_{ij},\widehat{\Delta}_{kl}}(u,v), \qquad \hat{\lambda}_{ij\Disp}=\frac{\hat{f}_{ij\Disp}}{\sqrt{C_{\Disp}}}
\end{align}
Similar remarks apply for other protected operators that can appear in the bOPE of the bulk conserved currents $J_\ell$.

\section{Derivation of the Ward identities}

\subsection{Ward identity for the displacement operator}\label{app:dispWard}
Here we review the derivation of \cite{Behan:2020nsf} of the displacement Ward identity. The starting point is the three-point function of the displacement operator $\Disp$ with the free bulk scalar, i.e.\footnote{The blocks for the bulk-bulk-boundary three-point functions of scalar operators were later derived using a group theoretical approach in \cite{Buric:2020zea}.}
\begin{align}\label{Displ3pt}
	\langle \phi(\vec{x}_1,y_1)\phi(\vec{x}_2,y_2)\Disp(\infty)\rangle= y_1 y_2 b_2^2 {}\hat{f}_{22\Disp}+b_1^2 {}\hat{f}_{11\Disp} \left[|\vec{x}_{12}|^2-(d-1) \left(y_1^2+y_2^2\right)\right]~.
\end{align}
We want to match this expression against the bulk OPE channel expansion, which receives a contribution from the $\phi^2$ as well as from the stress-tensor. The complete expression, i.e. including contributions from bulk descendants, is
\begin{align}\label{phiphiDblockexpa}
	\langle \phi(x_1)\phi(x_2)\Disp (\infty)\rangle &=b_{\phi^2 \Disp}\mathcal{W}_{\phi^2}^{\phi\phi\Disp}(\vec{x}_{12},y_1,y_2)+\frac{c_{\phi\phi T}}{C_T}\,{x}_{12}^\mu {x}_{12}^\nu\langle T_{\mu\nu}(x_2)\Disp(\infty)\rangle~.
\end{align}
The first term in the r.h.s. of the above equation is the $\langle\phi^2\Disp\rangle$ bulk block computed in \cite{Behan:2020nsf}
\begin{align}\label{Displphi2block}
	\mathcal{W}_{\phi^2}^{\phi\phi\Disp}(\vec{x}_{12},y_1,y_2)&=\frac{(d-1)(y_1+y_2)^2-|\vec{x}_{12}|^2}{4(d-1)}~.
\end{align}
The second term is the contribution from the bulk stress-tensor and reads \cite{McAvity:1993ue,McAvity:1995zd}
\begin{align}\label{TD2pt}
	\langle T_{\mu\nu}(x)\Disp(\infty)\rangle =&{b_{T\Disp}}{}\left(\delta_{\mu y}\delta_{\nu y}-\frac{1}{d}\delta_{\mu\nu}\right)~, \quad b_{T\Disp}=\frac{d\, C_{\Disp}}{d-1}~.
\end{align}
Note that bulk descendant operators of $T^{\mu\nu}$ do not enter into \eqref{phiphiDblockexpa}, since \eqref{TD2pt} is a constant. One can use further Ward identities for the displacement operator \cite{McAvity:1993ue,McAvity:1995zd} to relate the bOPE coefficient $b_{\phi^2 \Disp}$ to the one-point function of $\phi^2$:
\begin{align}\label{cphiphiT}
	b_{\phi^2 \Disp}=-a_{\phi^2}\,\, \frac{2^d (d-2)}{S_d}~, \quad S_d\equiv\text{Vol}(S^{d-1})=\frac{2 \pi ^{d/2}}{\Gamma \left(\frac{d}{2}\right)}~.
\end{align}
We can now equate \eqref{Displ3pt} to \eqref{phiphiDblockexpa} and solve for $\hat{f}_{11\Disp}$ and $\hat{f}_{22\Disp}$. The result is
\begin{align}\label{displRel2}
	{}\hat{f}_{11\Disp}=\frac{a_{\phi^2} 2^d C_T (d-2)-4  C_{\Disp}c_{\phi\phi T} S_d}{4C_T (d-1) S_d b_1^2}~,\qquad 
	{}\hat{f}_{22\Disp}=-\frac{a_{\phi^2} 2^d C_T (d-2)+4 C_{\Disp} c_{\phi\phi T} S_d}{2C_T S_d b_2^2}~.
\end{align}
The final formula \eqref{WardD20} is obtained by plugging into the above expression the values \eqref{cphiphiT2} of $c_{\phi\phi T}$ and $C_T$ corresponding to a $d$-dimensional free scalar field with unit normalization\footnote{In \cite{Osborn:1993cr} the free scalar $\bar\phi$ is canonically normalized and $ T_{\mu\nu}=\partial_\mu  {\bar\phi} \partial_\nu  {\bar\phi}+\dots$. When $\phi$ is unit-normalized,  we have $T_{\mu\nu} = {C_{\phi}}{}\partial_\mu \phi \partial_\nu \phi+\dots$ and the OPE coefficient $c_{\phi\phi T}$ is rescaled accordingly.}
\begin{align}\label{cphiphiT2}
	c_{\phi\phi T}= -\frac{d (d-2)}{2(d-1)S_d}~, \quad C_T= \frac{d}{(d-1) S_d^2}~.
\end{align}
The final result is
\begin{align}\label{displRelapp}
	{}\hat{f}_{11\Disp}=\frac{(d-2) \left(a_{\phi^2} 2^d+2 C_{\Disp} S_d^2\right)}{4 (d-1) S_d b_1^2}~, \quad 
	{}\hat{f}_{22\Disp}=\frac{(d-2) \left(2 C_{\Disp} S_d^2-a_{\phi^2} 2^d\right)}{2 S_d b_2^2}~.
\end{align}

\subsection{Ward identity for $\Disp_4^{(0)}$}\label{app:D4Ward}
The computations that lead to the Ward identity for the displacement operator can be easily adapted to derive the Ward identity satisfied by the special operators $\Disp_\ell^{(l)}$ of spin $l$ and scaling dimension $\widehat{\Delta}=d+\ell-2$ ($\ell$ is the spin of the bulk higher-spin current $J_\ell$, see section 2.2.2 of \cite{Behan:2020nsf} for details). Here we are going to derive the Ward identities for $\Disp_4^{(0)}$.

\subsubsection{Prelude: normalization of the bulk $J_4$}
In this section we consider the free scalar on $\mathbb{R}^d$, without a boundary. We find it convenient to work with canonical normalization, i.e. the propagator is
\begin{align}\label{two_pt_phi_homo}
	\langle \phi(x)\phi(0)\rangle= \frac{C_\phi}{|x|^{d-2}}~, \quad C_\phi = \frac{\Gamma \left(\frac{d }{2}-1\right)}{4 \pi ^{d /2}}~.
\end{align}
The spin 4 bulk current reads (for generic higher-spin currents, see e.g. \cite{Skvortsov:2015pea,Giombi:2016hkj, Giombi:2019enr})
\begin{equation}\label{spin4bulkdef}
	\begin{split}
		& J_{\mu \nu \rho \sigma}(x) =  {\mathcal N}_4 \bigg[ \phi\partial_{\mu} \partial_{\nu} \partial_{\rho} \partial_{\sigma}\phi  + \frac{12}{d-2} \delta_{(\rho\sigma} \partial_\mu \partial_{\nu)}\partial^\alpha\phi \partial_\alpha \phi+\frac{3(d+4)(d+2)}{d(d-2)}\partial_{(\mu}\partial_\nu \phi\partial_\rho\partial_{\sigma)}\phi \\
		&+  \frac{6}{d(d-2)}\delta_{(\mu\nu}\delta_{\rho\sigma)}\partial_\alpha\partial_\beta  \phi \partial^\alpha\partial^\beta  \phi -\frac{12(d+2)}{d(d-2)}\delta_{(\mu\nu} \partial^\alpha \partial_\sigma \phi \partial_{\rho)} \partial_\alpha  \phi-\frac{4(d+4)}{(d-2)}\partial_{(\mu}\phi \partial_{\nu} \partial_{\rho} \partial_{\sigma)}\phi \bigg]~,
	\end{split}
\end{equation}
where the brackets denote symmetrization of the indices (including the $1/4!$). The normalization ${\mathcal N}_4 $ of the current is fixed by the Ward identity. A convenient way to do it in practice is to pick a direction $y$ and fix the normalization of the commutator of $\phi$ with the charge operator supported on the hyperplane orthogonal to this direction, namely
\begin{align}
\begin{split}\label{spin4WardId}
& \int_{y>y'}\,\mathrm{d}^{d-1}\vec{x}'\, \langle J_{y\mu\nu\rho}(\vec{x},y)\phi(\vec{x}',y')\dots\rangle - \int_{y<y'}\,\mathrm{d}^{d-1}\vec{x}'\, \langle J_{y\mu\nu\rho}(\vec{x},y)\phi(\vec{x}',y')\dots\rangle \\
& = - \partial_{\mu}\partial_\nu \partial_{\rho}\langle \phi(\vec{x}',y')\dots\rangle~.
\end{split}
\end{align}
This gives
\begin{align}
	{\mathcal N}_4 = -\frac{d(d -2) }{8 (d +1) (d +3)}~.
\end{align}
Using the definition above we can compute e.g. 
\begin{align}
\begin{split}
& \langle J_4(x_1,\theta_1)J_4(x_2,\theta_2)\rangle = \frac{C_{J_4}}{{x_{12}}^{2d+4}}(\theta_1 \cdot I(x_{12})\cdot \theta_2)^4~,\\ &\langle \phi(x_1)\phi(x_2)J_4(\infty,\theta)\rangle = {c_{\phi\phi J_4}}{}({x}_{12}\cdot \theta)^4~,
\end{split}
\end{align}
to find
\begin{align}\label{J4twoandtree}
\begin{split}
	C_{J_4}=&192 C_{\phi}^2 (d +1) (d +2) (d +3) (d +4) ({\mathcal N}_4)^2=\frac{3 d ^2 (d +2) (d +4)}{(d +1) (d +3) S_d^2}~,\\
	 c_{\phi\phi J_4}=& C_{\phi}^2 (d -2) d  (d +2) (d +4){\mathcal N}_4 =-\frac{d ^2 (d +2) (d +4)}{8(d +1) (d +3) S_d^2}~.
\end{split}
\end{align}
As a check, the $C_{J_4}$ and $c_{\phi\phi J_4}$ given above reproduce the blocks coefficients of a spin 4 primary with  $\Delta = d+2$, exchanged in the GFF four-point function of $\phi$, when the latter is normalized as in eq.~\eqref{two_pt_phi_homo}. In particular, in $d$ dimensions the GFF coefficients are found to be \cite{Fitzpatrick:2012yx} \footnote{The result of eq. (11) in \cite{Fitzpatrick:2012yx} is multiplied by a factor of $2^\ell$ to match our conventions. The factor of $C_{\phi}^2$ on the l.h.s. above is due to the canonical normalization of the external scalar.}
\begin{align}
\begin{split}\label{MFTcoeff}
	&C_{\phi}^2P_{2\Delta_\phi+2n+\ell,\ell} \\ & \hspace{1 cm}=\frac{2^\ell  \left((-1)^\ell+1\right) \left(-\frac{d}{2}+\Delta _{\phi }+1\right)_n^2 \left(\Delta _{\phi }\right)_{\ell+n}^2}{\ell! n! \left(\frac{d}{2}+\ell\right)_n \left(-d+n+2 \Delta _{\phi }+1\right)_n \left(\ell+2 n+2 \Delta _{\phi }-1\right)_\ell \left(-\frac{d}{2}+\ell+n+2 \Delta _{\phi }\right)_n}~,
\end{split}
\end{align}
so that, using eq.~\eqref{J4twoandtree} we find precisely
\begin{align}
	\frac{c_{\phi\phi J_4}^2}{C_{J_4}}=C_{\phi}^2P_{d+2,4}~.
\end{align}

\subsubsection{Ward Identity with the boundary}


In the presence of the boundary, the non-conservation of $J_4$ in the directions orthogonal to the boundary is captured by 
\begin{align}
	\partial_\mu J^{\mu yyy}(\vec{x},y) = \delta(y) \Disp_4^{(0)} (\vec{x})~.
\end{align}
or equivalently 
\begin{align}\label{D4defPillow}
	J_4^{yyyy}(\vec{x},0)=-\Disp_4^{(0)}(\vec{x})~.
\end{align}
Therefore, after specifying the transverse indices in eq.~\eqref{spin4bulkdef} and using the Laplace equation to write $\lim_{y\rightarrow 0}\partial_y^2 \phi(\vec{x},y)= -\partial_a \partial_a \phi(\vec{x},0)$ we can find the explicit form of $\Disp_4^{(0)}$ for a free massless scalar in $d$ bulk dimensions
\begin{equation}\label{spin4Ddef}
	\begin{split}
		\Disp_4^{(0)}(\vec{x})&=  -{\mathcal N}_4 \bigg[\phi(\partial_a \partial_a)^2 \phi  - \frac{12}{d-2} \partial_\mu \phi \partial^\mu \partial_a \partial_a\phi+\frac{3(d+4)(d+2)}{d(d-2)}(\partial_a^2\phi) (\partial_b^2\phi)\\
		&+  \frac{6}{d(d-2)}\partial_\mu\partial_\nu  \phi \partial^\mu\partial^\nu  \phi -\frac{12(d+2)}{d(d-2)}\partial_\mu \partial_a \phi \partial^\mu \partial_a \phi+\frac{4(d+4)}{d-2}\partial_y\phi \partial_y\partial_a \partial_a\phi \bigg]_{y=0}~.
	\end{split}
\end{equation}
Note that in an interacting boundary condition this expression is understood as a non-singular OPE among the boundary modes $\phi$, $\partial_y\phi$ and their derivatives. 

Using this expression we can easily compute some data associated to $\Disp_4^{(0)}$ in the free boundary conditions, where correlators are simply given by Wick contractions. For instance we can compute
\begin{align}
	\langle \phi^2(\vec{x},y)\Disp_4^{(0)}(0)\rangle=\frac{b_{\phi^2\Disp_4^{(0)}}}{y^{d-2}(\vec{x}^2+y^2)^{d+2}}~,
\end{align}
to find
\begin{align}
b_{\phi^2\Disp_4^{(0)}}=-32 \,\kappa \,C_\phi^2 (d -2) d  \left(d ^2-1\right) {\mathcal N}_4 = \kappa\frac{4 d ^2 (d -1) }{(d +3) S_d^2}~,
\end{align}
where $\kappa=1$ $(-1)$ for the Neumann (Dirichlet, respectively) boundary condition. Similarly we can compute the two-point function of $\Disp_4^{(0)}$ in eq.~\eqref{D4twoptdef}, finding
\begin{align}\label{CD4free}
	\widehat{C}_{\Disp_4^{(0)}}=384 \,C_{\phi}^2 \,(d -1) (d +1)^2 (d +3) ({\mathcal N}_4)^2=\frac{6 d^2(d-1)}{(d+3)S_d^2}~.
\end{align}
Note that we get the same result for both Neumann and Dirichlet boundary condition.

More generally, for any interacting boundary condition, the following Ward identity holds
\begin{align}\label{WardD4phi2}
	\int\,\mathrm{d}^{d-1}\vec{x}'\, \langle \phi^2(\vec{x},y)\Disp_4^{(0)}(\vec{x}')\rangle=-\frac{1}{4}\partial_y^3\langle \phi^2(\vec{x},y)\rangle~.
\end{align}
This identity can be proven starting from eq.~\eqref{spin4WardId} (with two insertions of $\phi$), pulling out the integral contour to wrap the boundary and then taking the bulk OPE limit. The minus sign from eq.~\eqref{D4defPillow} is compensated by another minus sign which is due to a change in the orientation in the integration. One can also check the identity by an explicit computation in the Neumann and Dirichlet boundary conditions.

We now derive the identities satisfied by $\hat{f}_{ii\Disp_4^{(0)}}$ for interacting boundary conditions. For simplicity we go back to the CFT normalization, i.e. $\langle \phi\phi\rangle =1$. With this normalization eq.~\eqref{spin4WardId} is obeyed if we take the spin 4 current $J_4$ to be normalized as
\begin{align}
	{\mathcal N}_4 = -C_\phi\frac{d(d -2) }{8 (d +1) (d +3)}~,
\end{align}
where $C_\phi$ is the same quantity that appeared in eq.~\eqref{two_pt_phi_homo}. 
The starting point is the three-point function of the displacement operator $\Disp_4^{(0)}$ with the free bulk scalar. From eq.~(2.20) of \cite{Behan:2020nsf}, and after imposing the OPE relations, this reads
\begin{align}\label{Displ43pt}
\begin{split}
	&\langle \phi(\vec{x}_1,y_1)\phi(\vec{x}_2,y_2)\Disp_{4}^{(0)}(\infty)\rangle= y_1 y_2\,b_2^2 {}\hat{f}_{22\Disp_4^{(0)}}\left[|\vec{x}_{12}|^2-\frac{d-1}{3}(y_1^2+y_2^2)\right]\\
	&+ b_1^2 {}\hat{f}_{11\Disp_4^{(0)}} \left[|\vec{x}_{12}|^4-2(d+1)|\vec{x}_{12}|^2(y_1^2+y_2^2)+\frac{1}{3}(d^2-1) \left(y_1^4+6 y_1^2 y_2^2+y_2^4\right)\right]~.
\end{split}
\end{align}
We want to decompose the correlation function of eq.~\eqref{Displ43pt} in the bulk channel where only $\phi^2$ and $J_4$ can contribute, so we find
\begin{align}\label{phiphiD4blockexpa}
\begin{split}
	& \langle \phi(x_1)\phi(x_2)\Disp_4^{(0)}(\infty)\rangle \\ &=b_{\phi^2 \Disp_4^{(0)}}\left(\langle \phi^2(x_2)\Disp_4^{(0)}(\infty)\rangle+\dots\right)+\frac{c_{\phi\phi J_4}}{C_{J_4}}\,{x}_{12}^\mu {x}_{12}^\nu {x}_{12}^\rho{x}_{12}^\sigma\langle J_{\mu\nu\rho\sigma}(x_2)\Disp_4^{(0)}(\infty)\rangle~\\
	&=b_{\phi^2 \Disp_4^{(0)}}\mathcal{W}_{\phi^2}^{\phi\phi\Disp_4^{(0)}}(\vec{x}_{12},y_1,y_2)+\frac{c_{\phi\phi J_4}}{C_{J_4}}\,{x}_{12}^\mu {x}_{12}^\nu {x}_{12}^\rho{x}_{12}^\sigma\langle J_{\mu\nu\rho\sigma}(x_2)\Disp_4^{(0)}(\infty)\rangle~.
\end{split}
\end{align}
Note that, since $\langle J_4 (x)\Disp_{4}^{(0)} (\infty)\rangle$ is constant, bulk descendants of $J_4$ contribute trivially in the r.h.s. above. In the second line above, the first term is the $\langle\phi^2\Disp_4^{(0)}\rangle$ bulk block, which is computed by plugging $\widehat{\Delta}=d+2$ into eq.~(B.12) of \cite{Behan:2020nsf}
\begin{align}\label{Displ4phi2block}
	\mathcal{W}_{\phi^2}^{\phi\phi\Disp_4^{(0)}}(\vec{x}_{12},y_1,y_2)&=\frac{\left(d^2-1\right) (y_1+y_2)^4-6 (d+1) |\vec{x}_{12}|^2 (y_1+y_2)^2+3 |\vec{x}_{12}|^4}{16 \left(d^2-1\right)}~.
\end{align}
The other term is obtained using the results of \cite{Liendo:2012hy,Billo:2016cpy,Lauria:2018klo}
\begin{align}\label{J4D42pt}
\begin{split}
&	{x}_{12}^\mu {x}_{12}^\nu {x}_{12}^\rho{x}_{12}^\sigma\langle J_{\mu\nu\rho\sigma}(x_2)\Disp_4^{(0)}(\infty)\rangle \\  &\hspace{2cm} ={b_{J_4\Disp_4^{(0)}}}{}\frac{3 |\vec{x}_{12}|^4-6 (d+1) |\vec{x}_{12}|^2 (y_1-y_2)^2+(d^2-1) (y_1-y_2)^4}{(d+2) (d+4)}~.
\end{split}	
\end{align}

Putting all of this together, after comparing eq.~\eqref{phiphiD4blockexpa}-\eqref{Displ43pt}, we find that the bulk-to-boundary crossing symmetry for this correlator is satisfied only if
\begin{align}\label{WardD40raw}
\begin{split}
	b_1^2{}\hat{f}_{11\Disp_4^{(0)}}=& \frac{3 b_{\phi^2\Disp_4^{(0)}}}{16(d^2-1)}+\frac{3}{(d+2)(d+4)}\frac{c_{\phi\phi J_4}b_{J_4\Disp_4^{(0)}}}{C_{J_4}}~,\\
	b_2^2{}\hat{f}_{22\Disp_4^{(0)}}=& \frac{3 b_{\phi^2\Disp_4^{(0)}}}{4(1-d)}+\frac{12(d+1)}{(d+2)(d+4)}\frac{c_{\phi\phi J_4}b_{J_4\Disp_4^{(0)}}}{C_{J_4}}~.
\end{split}	
\end{align}
 Next, we are now going to discuss how to fix $b_{J_4\Disp_4^{(0)}}$ and $b_{\phi^2\Disp_4^{(0)}}$. First, it follows from the definition~\eqref{D4defPillow} and eq.~\eqref{J4D42pt} that
\begin{align}
	\langle J_{yyyy}(x_2)\Disp_4^{(0)}(\infty)\rangle =b_{J_4\Disp_4^{(0)}}\frac{(d-1)(d+1)}{(d+2) (d+4)}\equiv -\widehat{C}_{\Disp_4^{(0)}}~,
\end{align}
so that
\begin{align}\label{D4twopt}
	\widehat{C}_{\Disp_4^{(0)}}=-b_{J_4\Disp_4^{(0)}}\frac{(d-1) (d+1)}{(d+2)(d+4)}~.
\end{align}
Furthermore, we can use the Ward identity \eqref{WardD4phi2} in order to fix $b_{\phi^2\Disp_4^{(0)}}$ in terms of $a_{\phi^2}$. Upon performing the integral we find
\begin{align}\label{bJDWard}
		b_{\phi^2\Disp_4^{(0)}}=2^{d} C_\phi a_{\phi^2} \frac{d^2(d-1)(d -2)^2}{d+3}~.
\end{align}
Altogether, upon plugging equations~\eqref{D4twopt} and~\eqref{bJDWard} into eq.~\eqref{WardD40raw} we find
\begin{align}\label{WardD40rawfinal}
\begin{split}
	b_1^2{}\hat{f}_{11\Disp_4^{(0)}}=& \frac{3 a_{\phi^2} 2^{d-4} (d-2) d^2}{(d+1) (d+3) S_d}+\frac{(d-2) S_d\widehat{C}_{\Disp_4^{(0)}}}{8 (d^2-1)}~,\\
	b_2^2{}\hat{f}_{22\Disp_4^{(0)}}=& -\frac{3 a_{\phi^2} 2^{d-2} (d-2) d^2}{(d+3) S_d}+\frac{(d-2) S_d\widehat{C}_{\Disp_4^{(0)}}}{2 (d-1)}~.
\end{split}
\end{align}
Let us now study the decoupling limits where, from the expressions of $b_1$ and $b_2$ in eq.~\eqref{b1b2def} and assuming that $\hat{f}_{ii\Disp_4^{(0)}}$ remain finite in such limits, either the l.h.s. of the first equation above (for Dirichlet b.c.) or the l.h.s. of the second equation above (for Neumann b.c.) vanishes. Then, if we require that also the r.h.s. vanishes accordingly (as a sufficient condition for having finite block coefficients) we find that, at either N or D 
\begin{align}\label{CD4freeDec}
	a_{\phi^2}=\pm 2^{2-d}~,\quad \widehat{C}_{\Disp_4^{(0)}}=\frac{6 (d-1) d^2}{(d+3)S_d^2}~.
\end{align}
This result reproduces that of eq.~\eqref{CD4free}. At each decoupling limit the block coefficients become
\begin{align}
	a_{\phi^2}= &-2^{2-d}~,\quad \frac{(\hat{f}_{11\Disp_4^{(0)}})^2}{\widehat{C}_{\Disp_4^{(0)}}}=\text{indeterminate}~,\quad \frac{(\hat{f}_{22\Disp_4^{(0)}})^2}{\widehat{C}_{\Disp_4^{(0)}}}=\frac{3d^2}{2(d-1)(3+d)}~,\nonumber\\
	a_{\phi^2}= &2^{2-d}~,\quad \frac{(\hat{f}_{11\Disp_4^{(0)}})^2}{\widehat{C}_{\Disp_4^{(0)}}}=\frac{3d^2(d-2)^2}{32(d-1)(d+1)^2(3+d)}~,\quad \frac{(\hat{f}_{22\Disp_4^{(0)}})^2}{\widehat{C}_{\Disp_4^{(0)}}}=\text{indeterminate}.
\end{align}
As expected, these indeed reproduce the blocks coefficients of a $d-1$ dimensional GFF, respectively for the operator $\widehat{O}_2$ (of dimension $\widehat{\Delta}_2=d/2$) and for $\widehat{O}_1$ (of dimension $\widehat{\Delta}_1=d/2-1$). 

\section{More data of the minimal model b.c.}
\label{app:moredata}

Consider the minimal model b.c. of section~\ref{minimalbc}. At $g,h=0$ we can form two scalar boundary primaries of scaling dimension equal to four
\begin{align}
	\widehat{\mathcal{O}}(\vec{x})\equiv (\Phi_{(1,2)} \partial_{a}^2 \partial_y \phi)(x)~,\quad \widehat{\tau\bar{\tau}}(\vec{x}) \equiv \htau(x_1+i x_2)\bar{\htau}(x_1-i x_2)~.
\end{align}
These have tree-level two-point functions
\begin{align}
	\langle \widehat{\tau\bar{\tau}}(\vec{x})\widehat{\tau\bar{\tau}}(0)\rangle^{(0)} = \frac{\hC_{\widehat{\tau\bar{\tau}}}}{|\vec{x}|^8}~,\quad \langle \widehat{\tau\bar{\tau}}(\vec{x}) \widehat{\mathcal{O}}(0)\rangle^{(0)}  =0~,\quad \langle \widehat{\mathcal{O}}(\vec{x})\widehat{\mathcal{O}}(0)\rangle^{(0)}  = \frac{C_{\widehat{\mathcal{O}}}}{|\vec{x}|^8}~,
\end{align}
with $C_{\widehat{\tau\bar{\tau}}}=(\hC_{\htau}^{(0)})^2 =(1/4\pi^4)$ and $C_{\widehat{\mathcal{O}}} = 225/2\pi$. It will be convenient to work with a basis of unit-normalized operators, so we will define
\begin{align}
{S}_1 (\vec{x}) =\frac{1}{\sqrt{C_{\widehat{\tau\bar{\tau}}}}}\widehat{\tau\bar{\tau}}(\vec{x})~,\quad {S}_1 (\vec{x})=\frac{1}{\sqrt{C_{\widehat{\mathcal{O}}}}}\widehat{\mathcal{O}}(\vec{x})~.
\end{align}
At one-loop we have that
\begin{align}\label{matrixanomalous}
\begin{split}
	\delta \langle S_1 (\vec{x})S_1 (0) \rangle^{(1)}  & =-2\pi h\, C_{S_1 S_1, (1,3)}\frac{ \log x^2}{x^4}~,\\  \langle S_1 (\vec{x})S_2 (0) \rangle^{(1)}  & =-2\pi g C_{S_1 S_2, \Phi_{(1,2)}\partial_y \phi}\frac{ \log x^2}{x^4}~,\\
	 \delta \langle S_2 (\vec{x})S_2 (0) \rangle ^{(1)}  &=-2\pi h\, C_{S_2 S_2, (1,3)}\frac{ \log x^2}{x^4}~,
\end{split}
\end{align}
and it is easy to verify that\footnote{In our normalization, for the holomorphic three-point correlator with the stress-tensor we have that $\langle T(0)\phi(1)\phi(\infty)\rangle \propto \hat{\lambda}_{T\phi\phi} = -\frac{\hD_{\phi}}{2\pi}$.}
\begin{align}
	\hat{\lambda}_{S_1 S_1, (1,3)} =0~,\quad \hat{\lambda}_{S_1 S_2, \Phi_{(1,2)}\partial_y \phi} =\frac{3}{40 \sqrt{2 \pi }}~,\quad \hat{\lambda}_{S_2 S_2, (1,3)} =-\frac{\sqrt{3}}{2}~.
\end{align}
From the results in eq.~\eqref{matrixanomalous} the matrix of anomalous dimensions at the IR fixed point is
\begin{align}\label{MatrixAnomalousDim}
M\equiv \begin{pmatrix}
	0 & 2\pi g_* C_{S_1 S_2, \Phi_{(1,2)}\partial_y \phi}\\
	2\pi g_* C_{S_1 S_2, \Phi_{(1,2)}\partial_y \phi} & 2\pi h_*\, C_{S_2 S_2, (1,3)}
\end{pmatrix}~,
\end{align}
and it has eigenvalues
\begin{align}\label{gammapm}
\gamma_{\pm}^{(1)} = \frac{3}{20 m}\left(5\pm 3 \sqrt{3}\right)+O(1/m^2)~.
\end{align}
The corresponding unit-normalized eigenvectors are
\begin{align}
	V_{\pm} = \alpha_{\pm}^{(1)}S_1+\alpha_{\pm}^{(2)}S_2~,\quad \alpha_\pm^{(1)} = \frac{5\mp 3 \sqrt{3}}{\sqrt{6 \left(9\mp5 \sqrt{3}\right)}}, \quad \alpha_\pm^{(2)} = \frac{1}{\sqrt{3 \left(9\mp5 \sqrt{3}\right)}}.
\end{align}
 We now want to compute the OPE coefficients between the boundary modes of $\phi$ and any of the $\hat{\lambda}_{ij V_{\pm}}$ at the first non-trivial order, i.e. we want to compute the following three-point correlation functions
\begin{align}
	\langle \widehat{O}_i (\vec{x}_1) \widehat{O}_j (\vec{x}_2) V_{\pm}(\infty)\rangle.
\end{align}
For $\hat{\lambda}_{11 V_{\pm}}$ this is easily done using the modified Dirichlet boundary condition, so that we find
\begin{align}
\hat{\lambda}_{11 V_{\pm}} = \alpha_\pm^{(1)}/8+O(1/m)~.
\end{align}
For the other coefficients we simply invoke the exact relations of eq.~\eqref{ssspinconstr}, which we can exploit perturbatively i.e. by plugging $a_{\phi^2}=-1/2+ {8}/{m^2}+O(1/m^3)$ from eq.~\eqref{gammatauvsa_minimal_D} and $\Delta_{\pm}=4+\gamma_{\pm}^{(1)}+O(1/m^2)$ from eq.~\eqref{gammapm} and expanding to find that
\begin{align}
\hat{\lambda}_{12 V_{\pm}}=	-\frac{15 \alpha_\pm^{(1)} }{2 \sqrt{2}\left(5\pm3 \sqrt{3}\right)}+O(1/m)~,\quad \hat{\lambda}_{22 V_{\pm}}=	-\frac{9 \alpha_{\pm}^{(1)}}{m^2}+O(1/m^3)~.
\end{align}

\section{Parameters for rational approximations}\label{app:K123}
In the crossing equations that we need to approximate, the arbitrarily large variable $\hD$ appears in two places: the usual conformal blocks and the gamma functions of the exact relations. The greatest possible speed-up is achieved when both of these can be input to the numerical bootstrap as rational functions. To write a conformal block in this form, one must choose a single cutoff which we call $K_1$. For a block with spin $l$ and generic external dimensions, the algorithm in \cite{Kos:2014bka} produces a function with $\left \lfloor \frac{3K_1}{2} \right \rfloor + \mathrm{min}(K_1, l)$ poles. All of these are at or below the unitarity bound since they are associated with null descendants. As is also well known, the order of a given pole is at most $2$ with double poles only appearing when $\nu = \frac{d - 3}{2}$ is an integer.\footnote{Standard references define $\nu$ as $\frac{d - 2}{2}$. But here we keep the convention that $d$ is the number of \textit{bulk} dimensions.}

To approximate the exact relations using the steps in \cite{Behan:2020nsf}, two more cutoffs are required. One of them appears in
\begin{equation}
\frac{\Gamma \left ( \frac{d + l - \hD - 2}{2} \right )^2}{\Gamma \left ( \frac{d + l - \hD - 1}{2} \right )^2} \approx \frac{1}{K_2} \prod_{k = 0}^{K_2} \frac{(\hD - d - l - 2k + 1)^2}{(\hD - d - l - 2k + 2)^2}. \label{K2-def}
\end{equation}
This introduces $K_2 + 1$ double poles with the unusual property that they are above the unitarity bound. Some consequences of this are discussed in \cite{Behan:2020nsf,Behan:2018hfx}.

The other part that depends on a cutoff,
\begin{equation}
\frac{\Gamma \left ( \frac{\hD + l}{2} \right )^2}{\Gamma \left ( \frac{\hD + l + 1}{2} \right )^2} \approx \frac{(\frac{\hD + l + 1}{2})_{K_3}^2}{(\frac{\hD + l}{2})_{K_3 + 1}^2} \left [ \frac{2\hD + 2l + 4K_3 + 1}{4} + \frac{1}{8(2\hD + 2l + 4K_3 + 1)} \right ], \label{K3-def}
\end{equation}
has a more interesting effect on the degree of the approximation. This is because the numerator and denominator of \eqref{K3-def} both have zeros which can coincide with poles of the conformal blocks themselves.

The precise number of poles that \eqref{K3-def} can cancel or enhance depends on whether $\nu$ is an integer. Approximating the blocks using \texttt{PyCFTBoot}, as in \cite{Behan:2020nsf}, would limit us to non-integer values of $\nu$ resulting in some loss of precision. To avoid this, we have switched to a workflow where conformal block tables are first generated with \texttt{scalar\_blocks} \cite{scalarblocks}. Further numerical setup is then done in \texttt{PyCFTBoot} which can import tables from \texttt{scalar\_blocks}. A slight drawback to this approach is that we lose the ability to discard the poles whose residues vanish as a result of special dimension differences $\hD_{12}$ and $\hD_{34}$.\footnote{In large enough bootstrap problems, vanishing residues are moot because any block having special $\hD_{ij}$ will appear in at least one crossing equation with another block having generic $\hD_{ij}$. Nevertheless, this work only uses five crossing equations so there is still plenty of room to discard poles.} This has a bearing on all of the blocks in our setup (including the mixed ones) because $\hO_1$ and $\hO_2$ have scaling dimensions which differ by $1$. One needs to know whether this truncation has been done to properly account for \eqref{K3-def}.

\begin{table}
\centering
\begin{tabular}{c|c|c|c}
& \texttt{PyCFTBoot} & \texttt{scalar\_blocks}, $\nu \notin \mathbb{N}$ & \texttt{scalar\_blocks}, $\nu \in \mathbb{N}$ \\
\hline
$\#^{(2)}_>$ & $K_2 + 1$ & $K_2 + 1$ & $K_2 + 1$ \\
$\#^{(1)}_\leq$ & $\left \lfloor \frac{3K_1}{2} \right \rfloor + l - 2K_3 - 1$ & $\left \lfloor \frac{3K_1}{2} \right \rfloor + l - K_3 - 1$ & $\left \lceil \frac{K_1 + 1}{2} \right \rceil + 3l + 2\nu$ \\
$\#^{(2)}_\leq$ & $K_3 + 1$ & $0$ & $\left \lfloor \frac{K_1}{2} \right \rfloor - l - \nu - K_3 - 1$ \\
$\#^{(3)}_\leq$ & $0$ & $K_3 + 1$ & $0$ \\
$\#^{(4)}_\leq$ & $0$ & $0$ & $K_3 + 1$
\end{tabular}
\caption{A count of the number of poles from infinite families as a function of spin and the three cutoffs. As shown in \eqref{K3-def}, there is also one more pole below unitarity at $\hD = -l - 2K_3 - \frac{1}{2}$. This is a new simple pole for generic $\nu$ but it joins an existing pole when $\nu$ is a half-integer.}
\label{tab:poles}
\end{table}

All tolled, table \ref{tab:poles} describes the three most widely available approximations where $\#^{(n)}_>$ ($\#^{(n)}_\leq$) is the number of $n$'th order poles above (not above) the unitarity bound. For simplicity, we assume $l \leq K_1$ and $K_3 \leq \left \lfloor \frac{K_1 - 1}{2} \right \rfloor$. This work uses the last column and takes $(K_1, K_2, K_3) = (40, 12, 1)$.

\bibliography{bib}
\bibliographystyle{JHEP}

\end{document}